\documentclass[runningheads,a4paper]{article}

\usepackage{algorithm}
\usepackage{algorithmic}
\usepackage{graphicx}
\usepackage{floatflt}
\usepackage{wrapfig}
\usepackage{pstricks}
\usepackage{rotating}
\usepackage{subfigure}
\usepackage{amsmath}
\usepackage{amssymb}
\usepackage{theorem}
\usepackage[active]{srcltx} % for forward searching
\usepackage{amsmath,amssymb}
\usepackage{booktabs}
\usepackage{xcolor}
\usepackage{hyperref}
\usepackage{listings}
\usepackage[justification=RaggedRight, singlelinecheck=false]{caption}

\newcommand{\QED}{\hspace*{\fill}$\Box$}

\newtheorem{thm}{Theorem}
\newtheorem{lem}[thm]{Lemma}

\newtheorem{cor}[thm]{Corollary}

\newtheorem{rem}[thm]{Remark}
\newtheorem{ex}{Example}

\newenvironment{enumerate-} % low-profile enumerate
{\begin{enumerate}
    
   \setlength{\parskip}{-1ex}              % vertical space between paragraphs
   \setlength{\itemsep}{1.5ex}             % vertical space between items
}
{
 \end{enumerate}
}

\newcommand{\K}{{\mathcal K}}
\newcommand{\T}{{\mathcal T}}

\newcommand{\alg}{\mathcal} %'algebra' of given carrier set

\newcommand{\A}{\alg{A}}

\begin{document}

\title{On Invariant Synthesis for Parametric Systems}
% \titlerunning{On Invariant Synthesis for Parametric Systems}        % if too long for running head
% \authorrunning{Dennis Peuter, Viorica Sofronie-Stokkermans} % if too long for running
\author{Dennis Peuter   and 
        Viorica Sofronie-Stokkermans \\[3ex]
Universit{\"a}t Koblenz-Landau, Koblenz, Germany\\
              {$\{$dpeuter,sofronie$\}$}@uni-koblenz.de}    
%head

\maketitle

\begin{abstract}
We study possibilities for automated invariant 
generation in parametric systems. We use (a refinement of) an 
algorithm for symbol elimination in theory extensions to 
devise a method for iteratively strengthening certain 
classes of safety properties 
to obtain invariants of the system. We identify conditions under 
which the method is correct and complete, and situations in 
which the method is guaranteed to terminate. 
We illustrate the ideas on various examples.
\end{abstract}

\section{Introduction}

In the verification of parametric systems 
it is important to show that a certain property holds 
for all states reachable from the initial state.  
One way to solve such problems
is to identify an inductive invariant which entails the 
property to be proved. 
Finding suitable inductive invariants is
non-trivial -- the problem is 
undecidable in general; solutions have been proposed for 
specific cases, as discussed in what follows. 

\smallskip
\noindent 
In \cite{Kapur06a}, Kapur proposes methods for 
invariant generation in theories such as Presburger arithmetic, real
closed fields, and for polynomial equations and inequations with 
solutions in an algebraically
closed field. The main idea is
to use templates for the invariant (polynomials with
undetermined coefficients), and solve constraints for all paths and 
initial values to determine the coefficients. A similar idea was
used by Beyer et al. \cite{Rybalchenko} 
for constraints in linear real or rational arithmetic;
it is shown that if an invariant is
expressible with a given template, then it will be computed. 
Symbol elimination has been used for interpolation and invariant
generation in many papers. The methods proposed in \cite{Kapur06a}, 
where quantifier elimination or Gr{\"o}bner bases computation are used for 
symbol elimination, are one class of examples. Quantifier elimination
is also used by Dillig et al. in \cite{Dillig}.

\smallskip
\noindent 
 However, in some cases the investigated theories are complex (can be
extensions or combinations of theories) and do not allow quantifier
elimination. Methods for ``symbol elimination'' for such complex
theories have been proposed, in many cases 
in relationship with interpolant computation.  
In \cite{Yorsh05} Yorsh et al. studied interpolation in combinations
of theories; in \cite{Ghilardi14}, Brutomesso et al. extended these results to non-convex
theories. Interpolation in data structures by reduction was
analyzed by Kapur, Majumdar and Zarba in
\cite{Kapur06b}. Independently,  in \cite{Sofronie-ijcar06,Sofronie-lmcs-2008} 
Sofronie-Stokkermans analyzed possibilities of 
computing interpolants hierarchically, and in 
\cite{Sofronie-ijcar16,Sofronie-lmcs-2018} proposed a method of
hierarchical  symbol elimination which was used for interpolant
computation; already \cite{Sofronie-cade2013} mentions the possibility 
to infer constraints on parameters by hierarchical reasoning
followed by quantifier elimination.  

\smallskip
\noindent 
Symbol elimination 
can also be achieved 
using refinements of superposition. In \cite{BGW94}, Bachmair et al. mention 
the applicability of a form of hierarchical superposition to
second-order quantifier elimination (i.e. to symbol elimination). 
This idea and possible links to interpolation are also mentioned 
in Ganzinger et al. \cite{GSW04,GSW06}.  In \cite{Voronkov09}, Kov{\'a}cs
and Voronkov study inference systems and 
local derivations -- in the context of interpolant
generation -- and symbol elimination in proofs in such systems. 
The ideas are concretized using the superposition calculus and 
its extension LASCA (ground linear rational
arithmetic and uninterpreted functions). Applications to invariant
generation (briefly mentioned in \cite{Voronkov09}) are explored in detail 
in, among others, \cite{Voronkov09-fase,Voronkov-ijcar10}
-- there Vampire is used to generate a large set of invariants 
using symbol elimination; only invariants not implied by 
the theory axioms or by other invariants are kept 
(some of these tasks are undecidable). 
In \cite{Gleiss18}, Gleiss et al. analyze functional and temporal properties of loops. 
For this, extended expressions (introduced in \cite{Voronkov09-fase}) are used; 
symbol elimination {\`a} la \cite{Voronkov09} 
is used to synthesize invariants using quantification 
over iterations.

\smallskip
\noindent 
Various papers address the problem of strengthening a given formula 
to obtain an inductive invariant. 
In \cite{Bradley08}, Bradley proposes a goal-oriented invariant generation method for 
boolean/numeric transition systems,
relying on finding counterexamples. Such methods were implemented
in IC3 \cite{Bradley12}.  
For programs using only integers, Dillig et al.\ \cite{Dillig} use  
abductive reasoning based on quantifier elimination to obtain  
increasingly more precise approximations of inductive invariants
(termination is not guaranteed). 
In \cite{Kapur15}, Falke and Kapur analyze various ways of 
strengthening the formulae; depending upon how strengthening 
is attempted, their  
procedure may also determine whether 
the original formula is not an invariant. Situations in which 
termination is guaranteed are identified. 
In \cite{Shoham17}, Karbyshev et al.\ propose 
a method to generate universal invariants in
theories with the {\em finite model property} 
using diagram-based abstraction for invariant
strengthening; Padon et al.\ \cite{Shoham16}
identify sufficient conditions for the decidability of inferring 
inductive invariants in a given language ${\cal L}$ 
and also present undecidability results. 
Invariant synthesis for array-based systems is studied 
by Ghilardi et al.\ in \cite{Ghilardi10}; under local finiteness 
assumptions on the theory of elements 
and existence of well-quasi-orderings on configurations
termination 
is guaranteed.  
In \cite{Alberti14}, Alberti et al.\ use lazy abstraction with
interpolation-based refinement and discuss the applicability to
invariant synthesis. A system for verifying safety properties that are
``cubes'' and invariant generation in 
array-based systems  is described in \cite{Conchon12}. 
In \cite{Gurfinkel18}, Gurfinkel et al.\ propose an algorithm extending 
IC3 to support quantifiers for inferring universal
invariants 
in theories of arrays, combining quantified
generalizations (to construct invariants) with quantifier 
instantiation (to detect convergence).

\

\noindent {\bf Our contribution.} 
In this paper we continue the work on automated verification 
and synthesis in parametric systems 
started in \cite{Sofronie-tacas08,Sofronie-ijcar2010,Sofronie-cade2013}
by investigating possibilities 
for automated goal-oriented generation of inductive invariants. 
\noindent Our method starts with a universally quantified formula $\Psi$ and successively strengthens it, using a certain form of 
abductive reasoning based on symbol elimination. In case
of termination we prove that we obtain a universal 
inductive invariant that entails $\Psi$, or the answer ``no such
invariant exists''.  We identify
situations in which the method terminates. 
Our main results are: 
\begin{itemize}
\item We refine the symbol elimination method in theory
  extensions described in \cite{Sofronie-ijcar16,Sofronie-lmcs-2018} 
 (Sect.~\ref{qe-se}). This helps us obtain shorter formulae during invariant synthesis.
\item We propose a method for   goal-oriented synthesis of universally
  quantified invariants 
which uses symbol elimination in theory extensions (Sect.~\ref{verif-param-systems}). 
\item We identify conditions under which 
  our invariant generation method is partially correct
  (Sect.~\ref{verif-param-systems}) and  situations in which the
  method terminates (Sect.~\ref{termination}). 
\item We further refine the method (Sect.~\ref{refinements}) and provide examples in which the conditions we impose on the
  class of transition systems can be relaxed (Sect.~\ref{no-a5}).
\end{itemize}
The theories we analyze are extensions or 
combinations of theories and not required to have 
the finite model property --  required e.g.\ in 
\cite{Shoham17,Shoham16}. 
While we rely on methods similar 
up to a certain extent with the ones used in IC3 \cite{Bradley08,Bradley12},
and the ones in \cite{Ghilardi10,Kapur15,Shoham17}
we here use possibilities of complete instantiation in local theory 
extensions and exploit (and refine) efficient methods for symbol 
elimination in theory extensions proposed in
\cite{Sofronie-ijcar16,Sofronie-lmcs-2018}.

\medskip
\noindent 
{\bf Illustration.} Consider for instance the program in Fig.~\ref{fig2}, using subprograms
${\sf copy}(a, b)$,  which copies the array $b$ into array $a$, 
and ${\sf add1}(a)$, which adds 1 to every element of array $a$.

\begin{wrapfigure}[10]{l}{62mm}
\centering
\vspace{-3mm}
{\small 
\begin{verbatim}
  d1 = 1; d2 = 1; 
  copy(a, b); i:= 0; 
  while (nondet()) { 
     a = add1(a); 
     d1 = a[i]; d2 = a[i+1];
     i:= i + 1}
\end{verbatim}
}

\caption{Program using subprograms and global function updates}
\label{fig2}
\end{wrapfigure}

\noindent  
The task is to prove that if $b$ is an array with its
 elements sorted
in  increasing order then  the formula $\Psi := d_2 \geq d_1$ is an
invariant of the program. 
$\Psi$ holds in the initial state; it is an inductive invariant 
of the while
loop iff the formula

\medskip
\noindent $\begin{array}{@{}l} 
 \forall j (a'[j] = a[j]+1) \wedge d'_1 = a'[i] \wedge d'_2 = a'[i+1]
 \wedge  i' = i + 1 \\
\wedge ~d_1 \leq d_2 ~\wedge~ d'_1 >  d'_2 \\
\end{array}$  

\medskip
\noindent  is unsatisfiable. 
% It can be checked that this 
As this formula is satisfiable, $\Psi$ is
not an inductive invariant. 
 
\noindent 
We will show how to 
obtain the condition 
$\forall i (a[i] \leq a[i+1])$
which can be used to 
strengthen $\Psi := d_2 \geq d_1$
% . 
to the inductive invariant $(d_2 \geq d_1) \wedge
\forall i (a[i] \leq a[i+1])$. 

\

\noindent  While we rely on methods similar 
to the ones used in 
\cite{Bradley08,Ghilardi10,Dillig,Kapur15,Shoham16,Shoham17,Gurfinkel18}, 
there are several differences between our work and previous work.

\smallskip
\noindent 
The methods proposed in \cite{Kapur06a,Bradley08,Dillig,Kapur15}
cannot be used to tackle examples like the one in Fig.~\ref{fig2}: 
It is difficult to use templates in connection with additional
function symbols; in addition, the methods of
\cite{Bradley08,Dillig,Kapur15} 
can only handle numeric domains. 

\smallskip
\noindent 
The theories we analyze are typically extensions or 
combinations of theories and not required to have 
the finite model property --  which is required e.g.\ in 
\cite{Shoham17,Shoham16}. 

\smallskip
\noindent 
The method proposed  in \cite{Gurfinkel18} does not 
come with soundness, completeness and termination guarantees. 
We here use possibilities of complete instantiation in local theory 
extensions and exploit (and refine) the methods for symbol 
elimination in theory extensions proposed in
\cite{Sofronie-ijcar16,Sofronie-lmcs-2018}. 

\smallskip
\noindent 
The algorithm proposed in \cite{Ghilardi10}  
for theories of arrays uses a non-deterministic function 
{\sf ChooseCover} that returns a cover of a formula (as an approximation of 
the reachable states). If the theory of 
elements is locally finite it is proved that 
a universal formula $\Psi$ can be strengthened to a universal inductive invariant $I$ 
iff there exists a suitable {\sf ChooseCover} function for which the 
algorithm returns an inductive invariant strengthening $\Psi$.
In contrast, our algorithm is deterministic; we prove completeness under 
locality assumptions (holding if updates and properties are in the
array property fragment); our termination results are established for 
classes of formulae for which only finitely 
many atomic formulae formed with a fixed number of variables can be generated 
using quantifier elimination. 
In addition our method allows us to choose the language for the
candidate invariants (we can search for invariants not
containing certain constants or function symbols).

\smallskip
\noindent 
The methods in \cite{Voronkov09,Voronkov09-fase,Voronkov-ijcar10} use an 
approach different from ours: Whereas we start with a candidate 
invariant and successively strengthen it, there 
Vampire is used to generate a large set of invariants 
(by symbol elimination using versions of superposition combined with 
symbolic solving of recurrences);  completeness/termination are not guaranteed, 
although the method works well in practice.

\

\noindent {\em Structure of the paper.} In Section~\ref{prelim} we present the verification problems
we consider and the related reasoning problems, and present 
some results on 
local theory extensions. 
 In Section~\ref{qe-se} we 
present a 
method for symbol elimination in 
theory extensions introduced in
\cite{Sofronie-ijcar16,Sofronie-lmcs-2018} and propose an improvement of the method. 
In Section~\ref{verif-param-systems} 
we present an approach to invariant synthesis, and 
identify conditions under
which it is partially correct. 
Section~\ref{refinements} presents refinements 
 and a termination result.
In Section~\ref{implementation} we present the  tools we used for our tests and
the way we used them. Section~\ref{conclusion} contains conclusions and 
plans for future work.

\medskip
\noindent 
This paper is an extended version of \cite{Peuter-Sofronie-cade2019} containing full proofs and numerous examples.

\tableofcontents

\section{Preliminaries}
\label{prelim}

We consider one-sorted or many-sorted signatures $\Pi = (\Sigma, {\sf
  Pred})$, resp.\ $\Pi = (S, \Sigma, {\sf Pred})$,
where $S$ is a set of sorts, $\Sigma$ is a family of function symbols and ${\sf Pred}$
a family of predicate symbols. 
We assume known standard definitions from first-order logic  
(e.g.\ $\Pi$-structures, satisfiability, unsatisfiability, logical theories).  
We denote
``falsum'' with $\perp$. If $F$ and $G$ are formulae we 
write $F \models G$ (resp. $F \models_{\cal T} G$ -- also written as ${\cal T} \cup F \models G$) 
to express the fact that every model of $F$ (resp. every model of $F$ which is also a model of 
$\T$) is a model of $G$. $F \models \perp$ means that $F$ is
unsatisfiable; $F \models_{\T} \perp$ means that there is no model of
$\T$ in which $F$ is true.

\subsection{Verification problems for parametric systems}
% \label{verification}
\label{param-systems}

One of the application domains we  consider is 
the verification of parametric systems. 
For modeling such systems we use transition constraint systems 
$T = (\Sigma, {\sf Init}, {\sf Update})$ which specify:
the function symbols $\Sigma$ (including a set $V$ of functions with
arity 0 -- the ``variables'' of the systems) 
whose values change over time; 
a formula ${\sf Init}$ specifying the properties of initial states; 
a formula ${\sf Update}$
with 
function symbols in $\Sigma \cup \Sigma'$ 
(where $\Sigma'$ consists of copies of 
symbols in $\Sigma$, such that if $f \in \Sigma$ then $f' \in \Sigma'$
is the updated function 
after the transition). 
Such descriptions can be obtained from system specifications 
(for an example cf.\ \cite{Faber-Jacobs-Sofronie-07}). 
With every specification of a system $S$, a {\em background theory} ${\cal T}_S$ 
-- describing the data types used in the specification
and their properties -- is associated.

\smallskip
\noindent 
We can check in two steps whether a formula $\Psi$ is an inductive invariant of a 
transition constraint system 
$T {=} (\Sigma, {\sf Init}, {\sf Update})$, by checking whether: 
\begin{itemize}
\item[(1)] $ {\sf Init} \models_{\mathcal{T}_S} \Psi$; and 
\item[(2)] $\Psi,{\sf Update} \models_{\mathcal{T}_S} \Psi'$,  
where $\Psi'$ results from $\Psi$ by replacing 
% each $v \in V$ by $v'$ and 
each $f \in \Sigma$ by $f'$. 
\end{itemize}
\noindent 
Checking whether a formula $\Psi$ is an invariant can thus be reduced to
checking whether $\neg \Psi'$ is satisfiable or not w.r.t.\ a theory
$\T$. Even if $\Psi$ is a universally quantified formula (and thus
$\neg \Psi'$ is a ground formula) the theory $\T$ can be quite complex: 
it contains the axiomatization $\T_S$ of the datatypes used in the 
specification of the system, the formalization of the update rules, 
as well as the formula $\Psi$ itself.  
In \cite{Sofronie-tacas08,Sofronie-ijcar2010,Sofronie-cade2013} we
show that the theory $\T$ can often be expressed using a 
chain of extensions, typically including: 
$$ \T_0 ~~\subseteq~~ \T_1 = \T_0 \cup \Psi ~~\subseteq~~ \T = \T_0 \cup \Psi \cup {\sf
  Update}$$
with the property that checking satisfiability of ground formulae w.r.t.\ $\T$ can be
reduced to checking satisfiability w.r.t. $\T_1$ and ultimately to
checking satisfiability w.r.t.\ 
$\T_0$. This is the case for instance when the theory extensions in
the chain above are {\em local} (for definitions and further
properties cf.\ Section~\ref{sect:local}).

\medskip
\noindent 
Failure to prove (2) means that $\Psi$ is not an invariant 
or $\Psi$ is not inductive w.r.t.\ $T$. 
If $\Psi$ is not an inductive invariant, we can consider two orthogonal problems: 
\begin{enumerate}
\item[(a)] Determine constraints on 
  parameters which guarantee that $\Psi$ is an invariant. 
\item[(b)] Determine a 
formula  $I$ such that 
$\mathcal{T}_S \models I \rightarrow \Psi$ and 
$I$ is an inductive invariant.
\end{enumerate}

% \medskip
\noindent Problem (a) was studied in
\cite{Sofronie-ijcar2010,Sofronie-cade2013}.  
In \cite{Sofronie-ijcar16,Sofronie-lmcs-2018} we proposed a method for
hierarchical symbol elimination in theory extensions which allowed 
us to show that for local theory 
extensions the formulae obtained using this symbol elimination 
method are {\em weakest} constraints on parameters which 
guarantee that $\Psi$ is invariant. We present and improve 
this symbol 
elimination method in Section~\ref{qe-se}. 

\smallskip
\noindent 
In this paper we address
problem (b): in Section~\ref{sect:invgen} 
we use symbol elimination 
for giving a complete method for 
goal-oriented invariant generation, for invariants containing symbols
in a specified signature; we also identify some situations when
termination is guaranteed. 
The safety property and invariants we consider are conjunctions of
ground formulae and sets of  (implicitly universally
quantified) flat clauses of the form 
$\forall {\overline i}  (C_i({\overline c}, {\overline i}) \vee
C_v({\overline c}, {\overline i},
{\overline f}({\overline i})))$, where ${\overline c}$ are constants
or constant parameters, ${\overline f}$ are functional parameters, 
$C_i$ is a clause containing constants and universally quantified
variables, and $C_v$ a flat clause containing parameters, constants
and universally quantified variables.

\subsection{Local Theory Extensions}
\label{sect:local}

Let $\Pi_0 = (\Sigma_0, {\sf Pred})$ be a signature, and ${\cal T}_0$ be a 
``base'' theory with signature $\Pi_0$. 
We consider 
extensions $\T := {\cal T}_0 \cup \K$
of ${\cal T}_0$ with new function symbols $\Sigma$
({\em extension functions}) whose properties are axiomatized using 
a set $\K$ of 
clauses 
in the extended signature $\Pi = (\Sigma_0 \cup \Sigma, {\sf Pred})$, 
which contain function symbols in $\Sigma$. 
If $G$ is a finite set of ground $\Pi^C$-clauses\footnote{$\Pi^C$ is the
  extension of $\Pi$ with constants in a countable set $C$ of fresh
  constants.} and $\K$ a set of $\Pi$-clauses, we 
will denote by  ${\sf st}({\cal K}, G)$ (resp.\ ${\sf est}({\cal K}, G)$) the set of all 
ground terms (resp.\ extension ground terms, i.e.\ 
terms starting with a function in $\Sigma$) 
which occur in $G$ or ${\cal K}$.\footnote{We here regard every finite set $G$
of ground clauses as the ground formula $\bigwedge_{K \in G} K$.} 
If $T$ is a set of ground terms in the signature  $\Pi^C$, 
we denote by $\K[T]$ the set of all instances of $\K$ in which the terms 
starting with a function symbol in $\Sigma$ are in $T$. 
Let $\Psi$ be a map associating with 
every finite set $T$ of ground terms a finite set $\Psi(T)$ of ground terms. 
For any set $G$ of ground $\Pi^C$-clauses we write 
$\K[\Psi_{\cal K}(G)]$ for $\K[\Psi({\sf est}({\cal K}, G))]$.
We define: 
\begin{tabbing}
\= ${\sf (Loc}_f^\Psi)$~  \= For every finite set $G$ of ground clauses in $\Pi^C$ it holds that\\
\>\> $\T_0 \cup {\cal K} \cup G \models \bot$ if and only if $\T_0
\cup \K[\Psi_{\cal K}(G)] \cup G$ is unsatisfiable. 
\end{tabbing}
Extensions satisfying condition ${\sf (Loc}_f^\Psi)$ are called
{\em $\Psi$-local} \cite{Sofronie-tacas08,Ihlemann-Sofronie-ijcar10}. 
If $\Psi$ is the identity, i.e.\ $\K[\Psi_{\cal K}(G)]  = \K[G]$, 
we have a {\em local theory  
extension} \cite{Sofronie-cade05}.

\smallskip
\noindent {\bf Remark:} In \cite{Sofronie-tacas08,Ihlemann-Sofronie-ijcar10} we introduced and
studied  a notion of {\em extended locality}, in which the axioms in
$\K$ are of the form $\forall {\overline x} (\phi({\overline x})
\vee C)$, 
where $\phi$ is an arbitrary $\Sigma_0$-formula and $C$ a clause
containing extension symbols and the set $G$ contains ground formulae of the
form $\Psi \vee G_e$, 
where $\Psi$ is a  $\Sigma_0$-sentence and $G_e$ a ground clause containing 
extension symbols. While most of the results in this paper can be 
lifted by replacing ``locality'' with ``extended locality'', in this paper
we only refer to locality for the sake of simplicity.

\smallskip
\noindent 
\paragraph{{\bf Hierarchical reasoning in local theory extensions.}}
For ($\Psi$)-local theory extensions hierarchical reasoning is
possible. Below, we discuss the case of local theory extensions; 
similar results hold also for $\Psi$-local extensions. 
If ${\cal T}_0 \cup {\cal K}$ is a local extension of ${\cal T}_0$ and $G$ is a set of 
ground $\Pi^C$-clauses, 
then 
$ {\cal T}_0 \cup {\cal K} \cup G$ is unsatisfiable iff 
$ {\cal T}_0 \cup {\cal K}[G] \cup G$ is unsatisfiable. 
We can reduce this last satisfiability test to 
a satisfiability test w.r.t.\ ${\cal T}_0$ as follows:  
We purify ${\cal K}[G] \cup G$
by 
\begin{itemize}
\item[(i)] introducing (bottom-up) new constants $c_t$ for subterms $t = f(g_1, \dots, g_n)$ with $f \in \Sigma$, $g_i$ ground $\Sigma_0 \cup \Sigma_c$-terms,
\item[(ii)] replacing the terms $t$ with the constants $c_t$, and 
\item[(iii)] adding the definitions $c_t \approx t$ to a set $D$. 
\end{itemize}
 We denote by ${\cal K}_0 \cup 
G_0 \cup D$ the set of formulae obtained this way. 
Then $G$ is 
satisfiable w.r.t.\ ${\cal T}_0 \cup {\cal K}$ iff 
${\cal K}_0 \cup G_0 \cup {\sf Con}_0$ is satisfiable w.r.t.\
${\cal T}_0$, where 

\smallskip
\noindent ${\sf Con}_0  = \{ (c_1 \approx d_1 \wedge \dots \wedge c_n \approx d_n) \rightarrow c \approx d \mid 
  c \approx f(c_1, \dots, c_n), d \approx f(d_1, \dots, d_n) \in D \}.$ 

\begin{thm}[\cite{Sofronie-cade05}] 
\label{lemma-rel-transl}
If ${\cal T}_0 \subseteq {\cal T}_0 \cup {\cal K}$ is a 
local extension and $G$ is a finite set of
ground clauses, 
then we can reduce the problem of checking whether $G$ is 
satisfiable w.r.t.\ ${\cal T}_0 \cup {\cal K}$ to checking 
the satisfiability 
w.r.t.\ ${\cal T}_0$ of the formula ${\cal K}_0 \cup G_0 \cup {\sf
  Con}_0$ constructed as explained above. 

If ${\cal K}_0 \cup G_0 \cup {\sf Con}_0$ belongs to a decidable 
fragment of ${\cal T}_0$, we can use the decision procedure for 
this fragment to decide whether $ {\cal T}_0 \cup {\cal K} \cup G$
is unsatisfiable. 
\end{thm}
As the size of  
${\cal K}_0 {\cup} G_0 {\cup} {\sf Con}_0$ is polynomial in the size of $G$
(for a given ${\cal K}$), locality allows us to express the complexity 
of the ground satisfiability problem w.r.t.\ ${\cal T}_1$  
as a function of the complexity of the satisfiability 
of formulae w.r.t.\ ${\cal T}_0$.

\smallskip
% \subsubsection
\paragraph{{\bf Examples of local theory extensions}.}
\noindent ($\Psi$-)Local extensions can be 
recognized by showing that certain partial models embed into total
ones \cite{Ihlemann-Sofronie-ijcar10}. 
% This allowed us to identify  many classes of local theory extensions. 
Especially well-behaved are the theory extensions with property $({\sf
  Comp}_{f})$, stating that partial models can be made total without 
changing the universe of the model.\footnote{We use the index $f$ in 
$({\sf Comp}_{f})$ in order to emphasize that the property refers to 
completability of partial functions with a finite domain of definition.}
The link between embeddability 
% of partial models into total models 
and locality allowed us to identify  many classes of local theory
extensions: 
% Some examples are presented below. 

\begin{ex}[Extensions with free/monotone functions \cite{Sofronie-cade05,Sofronie-tacas08}] 
The following types of extensions of a theory $\T_0$ are local: 
\begin{enumerate}
\item Any extension of $\T_0$ with uninterpreted function
  symbols ($({\sf Comp}_f)$ holds). 
\item Any extension of a theory ${\cal T}_0$  for which $\leq$ is a partial order 
with functions monotone w.r.t.\ $\leq$ (condition $({\sf
  Comp}_f)$ holds if all models of $\T_0$ are complete lattices w.r.t.\ $\leq$). 
\end{enumerate}
\label{ex-monotone}
\end{ex}
\begin{ex}[Extensions with definitions \cite{JacobsKuncak,Sofronie-tacas08}]
Consider an extension of a theory $\T_0$ with a new function symbol $f$
defined by axioms of the form: 
%  case distinction using a set of clauses of the form

\smallskip
$~~~~~~~~{\sf Def}_f := \{ \forall {\overline x} (\phi_i({\overline x}) \rightarrow
    F_i(f({\overline x}), {\overline x})) \mid i =1, \dots, m \}$
    
\smallskip
\noindent (definition by ``case distinction'') where $\phi_i$ and $F_i$, $i = 1, \dots, m$, are formulae
over the signature of $\T_0$  such that the following hold: 
\begin{itemize}
\item[(a)] $\phi_i({\overline x}) \wedge \phi_j({\overline x})
  \models_{{\cal T}_0} \perp $  for $i {\neq} j$ and 
\item[(b)] 
${\cal T}_0 \models \forall {\overline x} (\phi_i({\overline x})
  \rightarrow \exists y (F_i(y, {\overline x})))$ for all $i \in \{ 1,
  \dots, m \}$.
\end{itemize}
Then the extension is local (and satisfies $({\sf Comp}_f)$). 
Examples: 
\begin{enumerate}
\item Any extension with a function $f$ defined by axioms of the form: 

\smallskip
$~~~~~~~~{\sf D}_f := \{ \forall {\overline x} (\phi_i({\overline x}) \rightarrow
 f({\overline x}) \approx t_i) \mid i = 1, \dots, n \}$   

\smallskip where
 $\phi_i$ are formulae over the signature of $\T_0$ such that (a) holds. 
\item Any extension of $\T_0 \in \{ {\sf LI}({\mathbb Q}), 
{\sf LI}({\mathbb R}) \}$ with functions satisfying 
axioms: 

\smallskip
$~~~~~~~~{\sf Bound}_f := \{ \forall {\overline x} (\phi_i({\overline x}) \rightarrow
 s_i \leq f({\overline x}) \leq t_i) \mid i = 1, \dots, n \}$

\smallskip
\noindent where $\phi_i$ are formulae over the signature of $\T_0$,
$s_i, t_i$ are $\T_0$-terms, condition (a) holds and $\models_{{\cal
    T}_0} \forall
{\overline x} (\phi_i({\overline x}) \rightarrow s_i \leq t_i)$
\cite{Sofronie-tacas08}.  
\end{enumerate}
\label{examples-local}
\end{ex}
\begin{ex}[The array property fragment \cite{manna-vmcai-06,Sofronie-tacas08}] 
In \cite{manna-vmcai-06} a decidable fragment of the theory of arrays
  is studied, namely the  {\em array  property fragment}.  
Arrays are regarded as functions with arguments of index sort 
and values of element sort.  
The index theory ${\cal T}_i$ is Presburger 
arithmetic. The element theory is parametric (for many-dimensional arrays: 
the element theories are parametric). 
Array property formulae are formulae of the form 
  $(\forall i)(\varphi_I(i) \rightarrow \varphi_V(i))$, where
  \begin{itemize}
\vspace{-1mm}
\item $\varphi_I$ (the index guard) 
is a positive Boolean combination of atoms of 
  the form $t {\leq} u$ or $t {=} u$ where $t$, $u$
  are either variables or ground terms of index sort;  
\item $\varphi_V$ (the value constraint) 
has  the property that any universally quantified 
   variable of index sort $i$ only occurs in a direct array read 
   $a(x)$ in $\varphi_V$ and array reads may not be nested.
\vspace{-1mm}
\end{itemize}
\noindent The \emph{array property fragment} consists of all existentially-closed
  Boolean combinations of quantifier-free
  formulae and array property formulae. 
 In \cite{manna-vmcai-06} it is shown that formulae in the array
property fragment have complete instantiation. 
In \cite{Sofronie-tacas08} we showed that  this  fragment 
satisfies a $\Psi$-locality
condition. % closure operator $\Psi$. 
\label{examples-apf-pointers}
\end{ex}

\section{Quantifier elimination and symbol elimination}
\label{qe-se}

We now present possibilities of symbol elimination in complex theories. 
  
\smallskip
\noindent A theory $\T$ over signature  
$\Pi$ {\em allows quantifier elimination} if for every formula $\phi$ over  
$\Pi$ there exists a quantifier-free formula $\phi^*$ over  
$\Pi$ which is equivalent to $\phi$ modulo $\T$. 
Examples of theories which allow quantifier elimination are
rational and real linear arithmetic (${\sf LI}({\mathbb Q})$, ${\sf
  LI}({\mathbb R})$), 
the theory of real closed fields, 
and the theory of absolutely-free data structures.

\smallskip
\noindent 
We first analyze possibilities of eliminating existential quantifiers
(EQE) in  combinations of disjoint theories.

\begin{rem}[EQE in combinations of disjoint theories] 
Let ${\cal T}_1$ and ${\cal T}_2$ be theories over disjoint
signatures $\Pi_1$ resp.\ $\Pi_2$. Let ${\cal T}$ be the two-sorted
combination of the theories ${\cal T}_1$ and ${\cal T}_2$, with 
signature $\Pi = (\{ s_1, s_2 \}, \Sigma_1 \cup \Sigma_2, {\sf Pred}_1
\cup {\sf Pred}_2)$, where every $n$-ary operation $f \in \Sigma_i$ has sort
$s_i^n \rightarrow s_i$, and every $m$-ary predicate symbol $p \in {\sf
  Pred}_i$ has arity $s_i^m$. 
Assume that ${\cal T}_1$ and ${\cal T}_2$  allow elimination of
existential quantifiers. 
Let $F(x, {\overline y})$ be a quantifier-free $\Pi$-formula.
 Then $\exists x F(x, {\overline y})$ is equivalent
w.r.t.\ ${\cal T}$ with a quantifier-free $\Pi$-formula $G({\overline
  y})$. 
\end{rem}

\noindent {\em Proof:} 
We can eliminate the existential variable $x$ from 
$\exists x F(x, {\overline y})$ by first bringing $F$ to disjunctive
normal form, $F(x, {\overline y})
  \equiv \bigvee_{i = 1}^n D_i(x, {\overline y})$, where every
  conjunction $D_i$ can be written as a 
conjunction $(D_i^1 \wedge D_i^2)$, where $D_i^j$ contains only atoms
over the signature $\Pi_j$, $j = 1,2$. 
Then 
$\exists x F(x, {\overline y}) \equiv \bigvee_{i = 1}^n \exists x
(D_i^1(x, {\overline y}) \wedge D_i^2(x, {\overline y}))$. 
If the variable $x$ is of sort $s_1$, then for every $i$, $x$ does not
occur in $D_i^2$ and 
$\exists x
(D_i^1(x, {\overline y}) \wedge D_i^2({\overline y})) \equiv \exists x
(D_i^1(x, {\overline y})) \wedge D_i^2({\overline y})$; a method for
quantifier elimination  for ${\cal T}_1$ can be used for computing
formulae ${\overline D}_i({\overline y})$ with 
$\exists x
(D_i^1(x, {\overline y})) \equiv {\overline D}_i({\overline y})$; the
case in which the sort of the variable $x$ is $s_2$ is similar. \QED

\subsection{Symbol elimination in theory extensions}
Let $\Pi_0 = (\Sigma_0, {\sf Pred})$. 
Let ${\cal T}_0$ be a $\Pi_0$-theory and $\Sigma_P$ be a set of parameters 
(function and constant symbols). 
Let $\Sigma$ be a signature such that $\Sigma \cap (\Sigma_0 \cup \Sigma_P) =
\emptyset$. %, containing functions not in $(\Sigma_0 \cup \Sigma_P)$. 
We consider the theory extension $\T_0 \subseteq \T_0 \cup \K$, where
${\cal K}$ is a set of clauses 
in the signature $\Pi = \Pi_0 \cup \Sigma_P \cup \Sigma$ in which all 
variables occur also below functions in $\Sigma_1 = \Sigma_P \cup
\Sigma$.  Consider the symbol elimination method in Algorithm~1
\cite{Sofronie-ijcar16,Sofronie-lmcs-2018}.

\begin{algorithm}[t]
\caption{Symbol elimination in theory extensions \cite{Sofronie-ijcar16,Sofronie-lmcs-2018}}
\label{algorithm-symb-elim}

\begin{description}
\item[Step 1] Let $\K_0 \cup G_0 \cup {\sf Con}_0$ be the set of
  $\Pi_0^C$-clauses obtained from $\K[T] \cup G$  after the purification step
  described in  Theorem~\ref{lemma-rel-transl} (with set of extension
  symbols $\Sigma_1$). 

\smallskip
\item[Step 2] Let $G_1 = {\mathcal K}_0 \cup G_0\cup {\sf Con}_0$. 
Among the constants in $G_1$, we identify 
\begin{enumerate}
\item[(i)] the constants
$c_f$, $f \in \Sigma_P$, where  $c_f$ is a constant
parameter or $c_f$ is 
introduced by a definition $c_f \approx f(c_1, \dots, c_k)$ in the hierarchical
reasoning method, % and 
\item[(ii)] all constants  ${\overline c_p}$ 
occurring as arguments of functions in $\Sigma_P$ in such definitions. 
\end{enumerate}
Let ${\overline  c}$ be the remaining constants. We replace the
constants in ${\overline  c}$
with existentially quantified variables ${\overline x}$, 
i.e.\ 
instead of $G_1({\overline c_p}, {\overline c_f}, {\overline c})$ 
we consider the formula $\exists {\overline x} G_1({\overline c_p},
{\overline c_f}, {\overline x})$.

\smallskip
\item[Step 3] Using a method for quantifier elimination in 
${\mathcal T}_0$
we can construct 
a formula  $\Gamma_1({\overline c_p}, {\overline c_f})$ equivalent to 
$\exists {\overline x} G_1({\overline c_p}, {\overline c_f},{\overline
  x})$
w.r.t.\ $\T_0$. %  (resp. $\T_0^*$). 

\smallskip
\item[Step 4] Let $\Gamma_2({\overline c_p})$ be the formula 
obtained by replacing back in $\Gamma_1({\overline c_p}, {\overline c_f})$ 
the constants $c_f$ introduced by definitions $c_f := f(c_1, \dots,
c_k)$ with the terms $f(c_1, \dots,c_k)$. We replace ${\overline c_p}$ with existentially quantified variables ${\overline y}$. 

\smallskip
\item[Step 5] Let $\forall {\overline y} \Gamma_T({\overline y})$ be
  $\forall {\overline y} \neg \Gamma_2({\overline y})$. 
\end{description}
\end{algorithm}

\begin{thm}[\cite{Sofronie-ijcar16,Sofronie-lmcs-2018}] 
Assume that ${\cal T}_0$ allows quantifier elimination.  
For every finite set of ground $\Pi^C$-clauses $G$, and every finite
set $T$ of ground terms over the 
signature $\Pi^C$ with ${\sf est}(G)
\subseteq T$, Steps 1--5 yield a universally quantified 
$\Pi_0 \cup \Sigma_P$-formula 
$\forall {\overline x} \Gamma_T({\overline x})$ 
such that ${\cal T}_0 \cup \forall {\overline y}
\Gamma_T({\overline y}) \cup {\cal K} \cup
  G$ is unsatisfiable. 
\label{inv-trans-qe}
\end{thm}

\begin{thm}[\cite{Sofronie-ijcar16,Sofronie-lmcs-2018}] 
   Assume that the theory extension ${\cal T}_0 \subseteq {\cal T}_0 \cup
  {\cal K}$ satisfies condition $({\sf Comp}_{f})$ and $\K$ is flat
  and linear and every variable occurs below an extension symbol. 
Let $G$ be a set of ground $\Pi^C$-clauses, 
and $\forall {\overline y} \Gamma_G({\overline y})$ 
be the formula obtained with Algorithm 1 for $T = {\sf est}(\K, G)$. 
Then $\forall y \Gamma_G(y)$ is entailed by every
universal formula $\Gamma$ with 
  ${\cal T}_0 \cup \Gamma \cup
  {\cal K} \cup G \models \perp$. 
\label{symb-elim-weakest}
\end{thm}
A  similar result holds if $T$ is the set of instances obtained from 
instantiation in a chain of theory extensions $\T_0 \subseteq \T_0
\cup \K_1 \subseteq \dots \subseteq \T_0
\cup \K_1 \cup \dots \cup \K_n$, all satisfying condition $({\sf
  Comp}_{f})$, where $\K_1, \dots, \K_n$ are all flat and linear and
every variable is  guarded by an extension function \cite{Sofronie-lmcs-2018}.

\medskip
\noindent {\bf Remark.} 
Algorithm 1 can be tuned 
to eliminate constants $c$ in a set $C_e$ which might occur as arguments to
parameters: All these constants, together with all
constants $c_f$ introduced by definitions $c_f = f(c_1, \dots, c_n)$ with 
some $c_i \in C_e$, are replaced with 
variables at the end of Step 2 and are eliminated  in Step 3.  

\subsection{An improved algorithm}
\noindent Quantifier elimination usually has high complexity
and leads to large formulae. Often, Algorithm~1 can be improved such that
QE is applied to smaller formulae: 

\begin{thm}
   Assume that $\K = \K_P \cup \K_1$ such that $\K_P$ contains only
   symbols in $\Sigma_0 \cup \Sigma_P$ and $\K_1$ is a set of
   $\Pi$-clauses such that
\[{\cal T}_0 \subseteq {\cal T}_0 \cup {\cal K}_P \subseteq {\cal T}_0
\cup {\cal K}_P \cup \K_1\] 
is a chain of theory extensions both satisfying condition $({\sf
  Comp}_{f})$ and having the property that all variables occur below an
extension function, and such that $\K$ is flat and linear. 
Let $G$ be a set of ground $\Pi^C$-clauses. 
Then the formula 
$\K_P \wedge 
\forall {\overline y} \Gamma_1({\overline y})$, 
where 
$\forall {\overline y} \Gamma_1({\overline  y})$ is obtained by 
applying Algorithm 1 to $\T_0 \cup \K_1 \cup G$, has the property 
that for every universal formula $\Gamma$ containing only parameters with 
  ${\cal T}_0 \cup (\K_P \cup
  \Gamma)  \cup G \models \perp$, we have 
$\K_P \wedge \Gamma  \models  \K_P \wedge \forall {\overline y} \Gamma_1({\overline y})$. 
\label{symb-elim-simplif}
\end{thm}
\noindent {\em Proof:} Assume that we have the following chain of
local theory extensions: 
$$ \T_0 \subseteq \T_0 \cup \K_P \subseteq \T_0 \cup \K_P \cup \K_1$$
We assume that the clauses in $\K_P \cup \K_1$ are flat and linear and 
that for each of these extensions any variable occurs below an
extension function. 
For the sake of simplicity we assume that $\Sigma$ contains only one
function symbol which we want to eliminate. For several function 
symbols the procedure is analogous. 
\noindent Let $G$ be a set of flat ground clauses. 
We know that the following are equivalent: 
\begin{itemize}
\item[(1)] $\T_0 \cup \K_P \cup \K_1 \cup G$ is satisfiable. 
\item[(2)] $\T_0 \cup \K_P \cup \K_1[G] \cup G$ is satisfiable,
  where the extension terms used in the instantiation correspond to 
 the set {\sf est}(G) = $\{ f(d_1), \dots, f(d_n) \}$, where for every
 $i$, $d_i = (d^1_i, \dots, d^{a(f)}_i)$.
\item[(3)] The set of formulae 
obtained after purification (in which
  each extension term $f(d)$ is replaced with a new constant $c_{fd}$,
  and we collect all definitions $c_{fd} \approx f(d)$ in a set ${\sf Def}$) and the
  inclusion of instances of the congruence axioms, 
$\T_0 \cup \K_P \cup (\K_1[G])_0
  \cup G_0 \cup {\sf Con}_0(G)$,  is satisfiable, where 
$$ {\sf Con}_0(G) = \{ (\bigwedge_{i = 1}^{a(f)} d^i_1 \approx d^i_2) \rightarrow  c_{fd_1} \approx c_{fd_2} \mid
f(d_1), f(d_2) \in {\sf est}(G) \}.$$ 
\item[(4)] $\T_0 \cup \K_P[G_1] \cup G_1$ is satisfiable, where $G_1 =
  (\K_1[G])_0
  \cup G_0 \cup {\sf Con}_0(G)$.
\item[(5)] The purified form of the formula above, $\T_0 \cup
  (\K_P[G_1])_0 \cup (G_1)_0 \cup {\sf Con}_0(G_1)$ is satisfiable. 
\end{itemize}
We use the following set of ground terms: 
$T = {\sf est}(G) \cup {\sf est}(G_1) = \{ f(d_1), \dots, f(d_n) \} \cup \{ g(c) \mid
g \in \Sigma_P, g(c) \in {\sf est}(G_1) \}$ (a set of flat terms, in
which we isolated the terms starting with the function symbol $f$). 
We apply Algorithm 1: 

\smallskip
\noindent {\bf Step 1:} We perform the hierarchical reduction in two steps. \\
In the first step we introduce a constant $c_{fd}$ for every term $f(d) \in {\sf
  est}(G)$. After this first reduction we obtain 
$G_1 = \K_1[G] \cup G_0 \cup {\sf Con}_0(G)$, where: 
$$\begin{array}{rcll}  
G_1 & = & G_0 \wedge & (\K_1[G])_0 \wedge
\displaystyle{\bigwedge_{f(d_1), f(d_2) \in {\sf est}(G)} \bigwedge_{i
    = 1}^{a(f)} d^i_1 \approx d^i_2 \rightarrow
c_{fd_1} \approx c_{fd_2}}
\end{array}$$
In the second reduction we replace every term 
$g(c) \in {\sf est}(G_1)$, $g \in \Sigma_P$, with a new constant $c_{gc}$ and 
add the corresponding congruence axioms and obtain:
$$G'_1 = (\K_P[G_1])_0 \wedge
(G_1)_0 \wedge {\sf Con}_0(G_1).$$

\noindent {\bf Step 2:} We want to eliminate the function symbol $f$. 
We assume that all the other
  function symbols are either parametric or in $\T_0$. 
We therefore replace the constants $c_{fd}$ with variables $x_{fd}$.  
 
\smallskip
\noindent {\bf Step 3:} 
Since $\K_P$ does not contain any function symbol in
$\Sigma$, neither $(\K_P[G_1])_0$ nor ${\sf Con}_0(G_1)$  
contain the variables $x_{fd}$. 
Therefore: 
\begin{eqnarray*}
\exists x_{fd_1}, \dots x_{fd_n} ~ G'_1(x_{fd_1}, \dots, x_{fd_n}) &
\equiv & (\K_P[G_1])_0 \wedge {\sf Con}_0(G_1)\wedge \exists x_{fd_1},
\dots x_{fd_n} (G_1)_0
\end{eqnarray*}
After quantifier elimination w.r.t.\ $\T_0$ we obtain a quantifier-free
$\Pi^C$-formula $D_0 \equiv \exists x_{fd_1},
\dots x_{fd_n} (G_1)_0$, hence 
$$\exists x_{fd_1}, \dots x_{fd_n} ~ G'_1(x_{fd_1}, \dots, x_{fd_n}) \equiv (\K_P[G_1])_0 \wedge {\sf Con}_0(G_1)\wedge D_0.$$ 
\noindent {\bf Step 4:} We replace back in the formula obtained this
way all constants $c_{gc}$, $g \in \Sigma_P, g(c) \in {\sf est}(G_1)$
with the terms $g(c)$ and remove ${\sf Con}_0(G_1)$.  This restores 
$\K_P[G_1]$ and all formulae which did not contain $f$ in $G'_1$. 
We obtain therefore: 
$$\Gamma_2({\overline d}) = \K_P[G_1]({\overline d}, {\overline c}, \overline{g(c)})
\wedge D({\overline d}, {\overline c}, \overline{g(c)})$$ 
We replace the constants ${\overline d}, {\overline c}$ with  variables ${\overline y}$ and obtain: 
$$\Gamma_2({\overline y}) = \K_P[G_1]({\overline y}, \overline{g(y)})
\wedge D({\overline y}, {\overline g(y)})$$ 
If we analyze the formula $\K_P[G_1]({\overline y}, {\overline
  g({\overline y})})$
obtained from $\K_P[G_1]$ by replacing the constants ${\overline d}, {\overline c}$ 
with  variables ${\overline y}$, we notice that the constants
substituted for variables in $\K_P[G_1]$ are replaced back with
variables. Thus, $\K_P[G_1]({\overline y}, \overline{g(y)}) =
\bigwedge_{i \in I} \K_P
\sigma_i$, a finite conjunction, where $\sigma_i : X \rightarrow X$ is
a substitution (not necessarily injective) that 
might rename the variables in $\K_P$. 

\smallskip
\noindent {\bf Step 5:} We negate $\exists {\overline y}
\Gamma_2({\overline y})$ and obtain:   

$$\neg \exists {\overline y}
\Gamma_2({\overline y}) \equiv \forall {\overline y} \neg
\Gamma_2({\overline y}) \equiv \forall {\overline y} (\bigvee_{i \in
  I} \neg \K_P \sigma_i
\vee \neg D({\overline y}, \overline{g(y)}))$$

\medskip

\noindent If the goal is to strengthen the already existing constraints $\K_P$ on the parameters
with a (universally quantified) additional condition $\Gamma$ such
that $\T_0 \cup \K_P \cup \Gamma \cup \K_1 \cup G \models \perp$ note
that:  
$$\begin{array}{llrl} 
\forall {\overline x} \K_P({\overline x}) \wedge \forall {\overline
  y} (\bigvee_{i \in I} \neg \K_P \sigma_i \vee \neg D({\overline y},
{\overline g({\overline y})})) &
  \equiv & \forall {\overline
    y} \big( & \!\!\!\! \bigvee_{i \in I} ({\forall {\overline x}} \K_P({\overline x}) \wedge \neg \K_P \sigma_i) 
  \vee \\
& & & ( {\forall {\overline x}} \K_P({\overline x}) \wedge \neg D({\overline y},
  {\overline g(y)}))  ~\big) \\
 & \equiv & \forall {\overline
    y} \big( &\!\!\!\! {\forall {\overline x}} \K_P({\overline x}) \wedge \neg D({\overline y},
  {\overline g({\overline y})}) ~\big) \\
& \equiv & {\forall {\overline x}} & \!\!\!\! \K_P({\overline x}) ~ \wedge ~ \forall
{\overline y} \neg D({\overline y},
  {\overline g({\overline y})})
\end{array}$$
We now prove that the formula  $\Gamma_1 = \forall {\overline y} \neg D({\overline y})$ is
the weakest formula with the property that $\T_0 \cup \K_P \cup \Gamma
\cup \K_1 \cup G$, i.e.\ 
that for every set $\Gamma$ of constraints on the parameters, 
if ${\mathcal T}_0 \cup \K_P \cup \Gamma \cup {\mathcal K}_1 \cup \dots \cup \K_n \cup
G$ is unsatisfiable then 
every model of $\T_0  \cup \K_P \cup \Gamma$ is a model of  $\T_0
\cup \K_P \cup \Gamma_1$. 

\smallskip
\noindent We know \cite{Ihlemann-Sofronie-ijcar10} that
if the extensions 
${\mathcal T}_0 \subseteq {\mathcal T}_0 \cup
  {\mathcal K}_P \subseteq \T_0 \cup \K_P \cup \K_1$ satisfy condition $({\sf Comp}_{f})$ then also the
  extensions 
${\mathcal  T}_0 \cup \Gamma \subseteq {\mathcal T}_0 \cup \Gamma \cup
{\mathcal K}_P \subseteq {\mathcal T}_0 \cup \Gamma \cup
{\mathcal K}_P \cup \K_1 $
  satisfy condition $({\sf Comp}_{f})$. If $\K_P \cup \K_1$ is flat and linear then the
  extensions are local. 
Let $T = {\sf est}(\K_1, G)$. 
By locality, ${\mathcal T}_0 \cup \Gamma \cup {\mathcal K}_P \cup \K_1 \cup G$ is
unsatisfiable
if and only if ${\mathcal T}_0 \cup \Gamma \cup {\mathcal K}_P \cup
\K_1[G] \cup G$
is unsatisfiable, if and only if  (with the notations in Steps 1--5) 
${\mathcal T}_0 \cup \Gamma \cup \K_P \cup ({\mathcal K}_1[G])_0 \cup G_0
\cup {\sf Con}_0 \cup {\sf Def}$ is unsatisfiable. 
Let ${\mathcal A}$ be a model of $\T_0  \cup \Gamma \cup \K_P$. Then ${\mathcal A}$ 
cannot be a model of $({\mathcal K}_1[G])_0 \cup G_0
\cup {\sf Con}_0 \cup {\sf Def}$, so (with the notation used when
describing Steps 1--5) there are no values for ${\overline d},
{\overline c}$ in the universe of ${\mathcal A}$ for which 
$D({\overline d}, {\overline c}, {\overline g}({\overline
  c}))$ would be true in ${\mathcal A}$, i.e.\ 
${\mathcal A} \not\models \exists {\overline y} D({\overline y}, {\overline g}({\overline
  y}))$. 
It follows that ${\mathcal A} \models \forall  {\overline y} \neg D({\overline y}, {\overline g}({\overline
  y}))$. Thus, ${\mathcal A} \models \K_P \wedge \Gamma_1$. 
\QED

\medskip
\noindent This improvement will be important for the method for invariant
generation we discuss in what follows. 
Further improvements are discussed in Section~\ref{refinements}. 

\subsection{Example}
\label{qe-example}

We illustrate the ideas of the improvement of the symbol elimination
algorithm based on Theorem~\ref{symb-elim-simplif} on an adaptation of an example first presented in
\cite{Sofronie-ijcar16,Sofronie-lmcs-2018}.

Let $\T_0 = {\sf LI}({\mathbb Q})$. 
Consider the extension of $\T_0$ with functions $\Sigma_1 =
\{ f, g, h, c \}$. Assume $\Sigma_P = \{ f, h, c \}$ and $\Sigma = \{
g \}$, and the properties of these function symbols 
are axiomatized by $\K_P \cup \K$, where 
\[ \begin{array}{lll} 
\K_P := \{ & \forall x, y (c < x \leq y \rightarrow f(x) \leq f(y)), & 
\forall x, y (x \leq y < c \rightarrow h(x) \leq h(y)) \quad \} \\
\K := \{ & \forall x (x \leq c \rightarrow g(x) \approx f(x)), ~~~~&  \forall x  ( c < x \rightarrow g(x) \approx h(x)) \quad \}.
\end{array} \] 
We are interested in generating a set of additional constraints on the functions $f$ and
$h$ which ensure that $g$ is monotone, e.g. satisfies 
$${\sf Mon}(g): ~~ \forall x, y (x \leq y \rightarrow g(x) \leq
g(y)),$$
i.e.\ a set $\Gamma$ of $\Sigma_0 \cup \Sigma_P$-constraints such that 
$\T_0 \cup \K_P \cup \Gamma \cup \K \cup \{ c_1 \leq c_2, g(c_1) > g(c_2) \}
\text{ is unsatisfiable,}$ 
where 
$G = \{ c_1 \leq c_2,  g(c_1) > g(c_2) \}$ is the Skolemized negation of ${\sf  Mon}(g)$.  
We have the following chain of theory extensions: 
\[ \T_0 \subseteq \T_0 \cup \K_P \subseteq \T_0 \cup \K_P \cup \K. \]  
Both extensions satisfy the 
condition ${\sf Comp}_f$, and 
$\T_0 \cup \K_P \cup \K \cup G$ is satisfiable iff
$\T_0  \cup \K_P \cup \K[G] \cup G$ is satisfiable, where: 
\[ \begin{array}{lll} 
\K[G] := \{ & c_1 \leq c \rightarrow g(c_1) \approx f(c_1), & c_2 \leq c \rightarrow g(c_2) \approx f(c_2), \\
& c < c_1 \rightarrow g(c_1) \approx h(c_1), & c < c_2 \rightarrow
g(c_2) \approx h(c_2) \quad \}.
\end{array} \] 
We construct $\Gamma$ as follows: 
\begin{description}
\item[Step 1] We perform a two-step reduction as we have a chain of two local extensions: First we
compute $\T_0 \cup \K[G] \cup G$, 
then purify it by 
introducing new constants $g_1, g_2$ for the terms $g(c_1), g(c_2)$. 
We obtain 

\smallskip
${\sf Def}_1 = \{ g_1 {\approx} g(c_1),  g_2 {\approx}
g(c_2) \}$ and 

\medskip
\noindent {$\begin{array}{ll} 
\K_0  \cup {\sf Con}_0 \cup G_0:= \{ & 
 c_1 \leq c \rightarrow g_1 \approx f(c_1) , ~~ c_2 \leq c \rightarrow g_2
 \approx f(c_2),  ~~\\
& c < c_1 \rightarrow g_1 \approx h(c_1),  ~~ c < c_2 \rightarrow g_2
\approx h(c_2), \\
& c_1 \approx c_2 \rightarrow g_1 \approx g_2,  ~ c_1 \leq c_2, ~~ g_1 > g_2 \quad \} 
\end{array}$ }

\medskip
In fact, we do not need the congruence axiom since $G \models c_1 \not\approx
c_2$. 

\medskip
In the next step, we introduce new constants $f_1,$ $f_2,$ $h_1,$ $h_2$  for the terms $f(c_1),$ $f(c_2),$
$h(c_1),$ and $h(c_2)$. 
We obtain:  

\smallskip
${\sf Def}_2 = \{ f_1 {=} f(c_1),
f_2 {\approx} f(c_2), h_1 {\approx} h(c_1), h_2 {\approx} h(c_2) \}$. 

\medskip
We do not need to effectively perform the instantiation of the axioms
in $\K_P$ or consider the instances of the congruence axioms for the
functions 
in $\Sigma_P$. 
We restrict to computing $(\K_0  \cup {\sf Con}_0 \cup  G_0)_0$: 

\medskip
\noindent {$\begin{array}{ll} 
(\K_0  \cup {\sf Con}_0 \cup G_0)_0:= \{ & 
 c_1 \leq c \rightarrow g_1 \approx f_1 , ~~ c_2 \leq c \rightarrow g_2
 \approx f_2,  ~~\\
& c < c_1 \rightarrow g_1 \approx h_1,  ~~ c < c_2 \rightarrow g_2
\approx h_2, \quad \\
& c_1 \leq c_2, ~~ g_1 > g_2 ~ \} 
\end{array}$ }

\medskip
\item[Step 2] The parameters are contained in the set $\Sigma_P = \{
  f, h, c \}$.  
We want to eliminate the function symbol $g$, so
  we replace $g_1, g_2$ with existentially
  quantified variables $z_1, z_2$ and obtain an existentially
  quantified formula 

\medskip
$\begin{array}{ll} 
\exists z_1,
  z_2 & [(c_1 \leq c \rightarrow z_1 \approx f_1) \wedge (c_2 \leq c \rightarrow z_2
 \approx f_2) \wedge \\
& (c < c_1 \rightarrow z_1 \approx h_1) \wedge (c < c_2 \rightarrow z_2
\approx h_2) \wedge  \\
& c_1 \leq c_2 \wedge  z_1 > z_2]
\end{array}$

\medskip
\item[Step 3] If we use QEPCAD for quantifier elimination we obtain the
  formula: \\
$\Gamma_1(c_1, c_2, c, f_1, f_2, h_1, h_2)$:  

\smallskip
$ \begin{array}{ll} 
c_2 > c_1 \wedge \big(  & ( c_2 \leq c ~\wedge~ f_2 < f_1 ) ~\lor ~ \\
& ( c_1 > c ~\wedge~ h_2 < h_1 ) ~\lor~ \\
& ( c_1 \leq c ~\wedge~ c_2 > c ~\wedge~ h_2 < f_1 )~ \big)
\end{array}
$

\medskip
\item[Step 4] We construct the formula $\Gamma_2(c_1, c_2, c)$ 
from $\Gamma_1$ 
% (c_1, c_2, c, f_1,  f_2, h_1, h_2)$ 
by replacing $f_i$ by $f(c_i)$ and $h_i$ by
  $h(c_i)$, $i = 1,2$. We obtain: 

\medskip
$ \begin{array}{ll} 
c_2 > c_1 \wedge \big( ~ & ( c_2 \leq c ~\wedge~ f(c_2) < f(c_1) )
~\lor~ ( c_1 > c ~\wedge~ h(c_2) < h(c_1) ) ~\lor \\
& ( c_1 \leq c ~\wedge~ c_2 > c ~\wedge~ h(c_2) < f(c_1) )~ \big)
\end{array}$

\medskip
\noindent After replacing $c_1$ with a variable $x_1$ and $c_2$ with a variable
$x_2$ we obtain: 

\medskip
$ \begin{array}{ll} 
x_2 > x_1 \wedge \big( ~ & ( x_2 \leq c ~\wedge~ f(x_2) < f(x_1) )
~\lor~  ( x_1 > c ~\wedge~ h(x_2) < h(x_1) ) ~\lor \\
& ( x_1 \leq c ~\wedge~ x_2 > c ~\wedge~ h(x_2) < f(x_1) ) ~ \big)
\end{array}$

\medskip
\item[Step 5] After negation and universal quantification of the
  variables we obtain: 
$$ \begin{array}{lll} 
& \forall x_1, x_2 \big(x_2 \leq x_1 \lor \big( & ( x_2 \leq c \rightarrow f(x_1) \leq f(x_2) ) ~\wedge \\
& & ( x_1 > c \rightarrow h(x_1) \leq h(x_2) ) ~\wedge\\
& & ( x_1 \leq c ~\wedge~ x_2 > c \rightarrow f(x_1) \leq h(x_2) ) ~
\big)\big)\\
& &  \\
 \text{ or equivalently:} & 
\forall x_1, x_2 \big(x_1 < x_2 \rightarrow \big( & ( x_2 \leq c \rightarrow f(x_1) \leq f(x_2) ) ~\wedge \\
& & ( x_1 > c \rightarrow h(x_1) \leq h(x_2) ) ~\wedge\\
& & ( x_1 \leq c ~\wedge~ x_2 > c \rightarrow f(x_1) \leq h(x_2) ) ~ \big)\big)
\end{array}$$
\end{description}

\

\section{Goal-oriented invariant synthesis} 
\label{verif-param-systems}

\smallskip
\noindent Let $S$ be a system, $\T_S$ be the theory and 
$T {=} (\Sigma_S, {\sf Init}, {\sf Update})$ the transition
constraint system associated with $S$. 
We assume that $\Sigma_S = \Sigma_0 \cup \Sigma_P \cup \Sigma$, where
$\Sigma_0$ is the signature of a ``base'' theory $\T_0$, $\Sigma_P$ is
a set of function symbols assumed to be parametric, and $\Sigma$ is a
set of functions (non-parametric) disjoint from $\Sigma_0 \cup
\Sigma_P$. We do not restrict to arithmetic as in \cite{Dillig} or to theories with the
finite model property as in \cite{Shoham16}. The theories we consider
can include for instance 
arithmetic and thus may have infinite models.

\smallskip
\noindent In this paper we consider only transition systems $T {=} (\Sigma_S,
{\sf Init}, {\sf Update})$ for which ${\sf Init}$ is a universal formula describing the
initial states and ${\sf Update}$ is a universal formula describing 
(possibly global) updates of functions in a set $F
\subseteq \Sigma$. Variable updates are a special case of updates (recall that variables
  are 0-ary functions). Since ground formulae are,
  in particular, also universal formulae, we allow in particular also 
initial conditions or updates expressed as ground formulae.

\medskip
\noindent We consider universal formulae $\Psi$ which are 
conjunctions of clauses of the form 
$\forall {\overline x} (C_i({\overline x}) \vee C_v({\overline x}, {\overline f}({\overline x}))$, 
where $C_i$ is a $\T_0$-clause and $C_v$ a flat clause over $\Sigma_0
\cup \Sigma_P$.\footnote{We use the following abbreviations: ${\overline x}$ for $x_1, \dots,
x_n$; ${\overline f}({\overline x})$ for $f_1({\overline x}), \dots,
f_n({\overline x})$.} 
Such formulae describe ``global'' properties of the function 
symbols in $\Sigma_P$ at a given moment in time, e.g.\ equality
of two functions (possibly representing arrays), or monotonicity of a
function. They can also describe properties of individual elements
(ground formulae are considered to be in particular universal
formulae).

\medskip
\noindent If the formula $\Psi$ is not an inductive invariant, 
our goal is to obtain a universally quantified inductive invariant
$I$ in a specified language 
(if such an inductive invariant exists) such that $I \models_{\T_S}
\Psi$, or a proof that there is no universal inductive
invariant that entails $\Psi$.

\

\noindent We make the following assumptions: 
Let ${\sf LocSafe}$ be a class of universal formulae over $\Sigma_S$. 
\begin{description}
\item[(A1)] There exists a chain of local theory extensions ${\cal
    T}_0 \subseteq \dots \subseteq \T_S \cup {\sf Init}$ such that in
  each extension all variables occur below an extension function.
\item[(A2)] For every $\Psi \in {\sf LocSafe}$ 
there exists a chain of local theory extensions
${\cal T}_0 \subseteq \dots \subseteq \T_S \cup \Psi$ such that in
  each extension all variables occur below an extension function.
\item[(A3)] 
${\sf Update} = \{ {\sf Update}_f \mid f \in F \}$ consists of update
axioms for functions in a set $F$,  where, for every $f
\in F$, ${\sf Update}_f$ has the form 
${\sf Def}_f := \{\forall {\overline x} (\phi^f_i({\overline x}) \rightarrow
C^f_i({\overline x}, f'({\overline x})))\mid i \in I \}$,
such that (i) 
$\phi_i({\overline x}) \wedge \phi_j({\overline x})
  \models_{{\cal T}_S} \perp $  for $i {\neq} j$, 
(ii) ${\cal T}_S \models
  \bigvee_{i = 1}^n \phi_i$,  and (iii) $C^f_i$ are conjunctions of
  literals and 
${\cal T}_S \models \forall {\overline x} (\phi_i({\overline x})
  \rightarrow \exists y (C^f_i({\overline x}, y)))$ for all $i \in I$. 
\footnote{The update axioms 
describe the change of the functions in a set  $F \subseteq \Sigma$, 
depending on a finite set $\{ \phi_i \mid i \in I \}$ of 
mutually exclusive conditions over non-primed symbols. \\
In particular we can consider definition updates of the form
${\sf D}_{f'}$ 
or updates of the form ${\sf Bound}_{f'}$ 
 as discussed in Example~\ref{examples-local}.} 
\end{description}
In what follows, for every formula $\phi$ containing 
% variables in $V$ and 
function symbols in $\Sigma$ we denote by $\phi'$ the
formula obtained from $\phi$ by replacing 
% every variable $v \in V$ with $v'$ and 
every function symbol $f \in \Sigma$ with the corresponding symbol $f'
\in \Sigma'$. 
\begin{thm}[{\small \cite{Sofronie-tacas08,Sofronie-ijcar2010}}] 
The following hold under assumptions ${\bf (A1)}-{\bf (A3)}$: 
\begin{itemize}
\item[(1)]  If ground satisfiability w.r.t.\ ${\cal T}_0$ is decidable,
  then the problem of checking whether a formula $\Psi \in {\sf LocSafe}$ 
is an inductive invariant of $S$ is decidable. 
\item[(2)] If ${\cal T}_0$ allows quantifier elimination and 
the initial states or the updates contain parameters, the 
symbol elimination method in Algorithm 1 yields constraints on these parameters that guarantee that $\Psi$ is
an inductive invariant. 
\end{itemize}
\label{thm-dec-inv-checking-synthesis}
\end{thm} 

\label{sect:invgen}
\noindent We now study the problem of inferring -- in a goal-oriented 
way -- universally
quantified inductive invariants. 
The method we propose 
is described in Algorithm~\ref{fig-inv-gen}.

\begin{algorithm}[t]
\caption{Successively strengthening a formula to an inductive  invariant} 
\label{fig-inv-gen}
\begin{tabular}{ll} 
{\bf Input:} & $T = (\Sigma_S, {\sf Init}, {\sf Update})$ transition system; $\Sigma_P \subseteq \Sigma_S$; 
%% satisfying conditions {\bf (A1),(A2),(A3),(A4)}  \\
$\Psi \in {\sf LocSafe}$, formula over $\Sigma_P$ ~~~~~\\
{\bf Output:} & 
Inductive invariant $I$ of $T$ that entails $\Psi$ and contains only function symbols in $\Sigma_P$\\
&  (if such an invariant exists). \\
\hline 
\end{tabular} 

\begin{tabular}{l} 
1: $I := \Psi$ \\
2: {\bf while} $I$ is not an inductive invariant for $T$ {\bf do:} \\
$~~$ {\bf if} ${\sf Init} \not\models I$ {\bf then} {\bf return} ``no
universal inductive invariant over $\Sigma_P$ entails $\Psi$'' \\
%% $~~$ {\bf if} $I$ is not invariant under updates {\bf do:} \\
$~~$ {\bf if} $I$ is not preserved under ${\sf Update}$ {\bf then} Let
$\Gamma$ be obtained by eliminating \\
$~~$ all primed variables and symbols not in $\Sigma_P$ from $I \wedge {\sf
  Update} \wedge \neg I'$; \\
$~~$ $I := I \wedge \Gamma$\\
3: {\bf return} $I$ is an inductive invariant 
\end{tabular} 
\end{algorithm}

\medskip
\noindent 
In addition to assumptions {\bf (A1), (A2), (A3)} we now consider the
following assumptions (where $\T_0$ is the base theory in assumptions {\bf (A1)--(A3)}): 
\begin{description}
\vspace{-1mm}
\item[(A4)] Ground satisfiability in ${\cal T}_0$ is decidable; 
  ${\cal T}_0$ allows quantifier elimination. 
\item[(A5)] All candidate invariants $I$ computed in the while loop in 
  Algorithm~\ref{fig-inv-gen} are in
  ${\sf LocSafe}$, and all local extensions in ${\sf LocSafe}$ satisfy
  condition $({\sf Comp}_f)$.
\vspace{-1mm}
\end{description}
We first prove that under assumptions $({\bf A1})-({\bf A5})$ 
the algorithm is partially correct
(Theorem~\ref{inv-gen-correctness}). 
Then we identify conditions under which the locality assumption
({\bf A5}) holds, so does not have to be stated explicitly (Section~\ref{no-a5}), 
and conditions under which the algorithm terminates (Section~\ref{termination}). 
\begin{lem}
If Algorithm~\ref{fig-inv-gen} terminates and returns a formula $I$, then $I$ is an
invariant of $T$ containing only function symbols in $\Sigma_P$ that entails $\Psi$. 
\end{lem}
{\em Proof:} Follows from the loop condition. \QED

\begin{lem} Under assumptions {\bf (A1)}--{\bf (A5)},  
if there exists a universal inductive invariant $J$
containing only function symbols in $\Sigma_P$ that
entails $\Psi$, then 
$J$ entails every candidate invariant $I$ generated in the while
loop of Algorithm~\ref{fig-inv-gen}. 
\label{lemma-entails}
\end{lem}

\noindent {\em Proof:} Proof by induction on the number of iterations in which
the candidate invariant $I$ is obtained. 
\noindent If $i = 1$, then $I_1 = \Psi$, hence $J \models \Psi =
I_1$.

\noindent Assume that the property holds for the candidate invariant $I_n$
generated in $n$ steps. Let $I_{n+1}$ be generated in step $n+1$. 
In this case there exist candidate invariants $I_1, \dots, I_n$ 
containing only function symbols in $\Sigma_P$ such that: 
(i) $I_1 = \Psi$; (ii) for all $1 \leq i \leq n$, ${\sf Init} \models I_i$; 
(iii) for all $1 \leq i \leq n$, $I_i$ is not an inductive invariant,
  i.e.\ $I_i \wedge {\sf Update} \wedge \neg
  I'_i$ is satisfiable and $\Gamma_i$ is obtained by eliminating the
  primed function symbols and all function symbols not in $\Sigma_P$; 
(iv) for all $1 \leq i \leq n$, $I_{i+1} = I_i \wedge \Gamma_i$.

We prove that $J \models_{{\cal T}_S} I_{n+1}$, i.e.\ that  
$J \models_{{\cal T}_S} I_{n} \wedge \Gamma_n$. 
By the induction hypothesis, $J \models_{{\cal T}_S} I_n$, hence 
$J \equiv_{{\cal T}_S} J \wedge I_n$. 
We know that $J$ is an inductive invariant, i.e.\ 
$J \wedge {\sf Update} \wedge \neg J'$ is
unsatisfiable. Therefore 
$ (J \wedge I_n) \wedge {\sf Update} \wedge (\neg J'
\vee \neg I'_n)$ is unsatisfiable, 
hence, in particular,
$ J \wedge I_n \wedge {\sf Update} \wedge \neg I'_n \text{ is unsatisfiable.}$
By Theorem~\ref{symb-elim-weakest}, the way $\Gamma_n$ is
constructed, and the fact that $J$ is a universal formula containing
only function symbols in $\Sigma_P$, 
we know that  $J \models_{{\cal T}_S} \Gamma_n$. By the induction
hypothesis, $J \models_{{\cal T}_S} I_n$.  
Thus, $J \models_{{\cal T}_S} I_n \wedge \Gamma_n$, so $J \models_{{\cal T}_S} I_{n+1}$. This
completes the proof. \QED

\begin{thm}[Partial Correctness]
Under assumptions {\bf (A1)}--{\bf (A5)}, 
if Algorithm~\ref{fig-inv-gen} terminates, then its output
is correct.
\label{inv-gen-correctness}
\end{thm}
{\em Proof (Sketch):} If Algorithm~2 terminates with output $I$, 
then the condition of the while loop must be false for $I$, so $I$ is an
invariant. 
Assume that Algorithm~2 terminates because ${\sf Init}
\not\models_{{\cal T}_S} I$ returning ``no universal inductive invariant 
over $\Sigma_P$ entails $\Psi$''. Then there exists a model $\A$ of ${\sf
  Init}$ and $\T_S$ which is not a model of $I$. 
Assume that there exists a universal inductive invariant $J$ over
$\Sigma_P$ that entails $\Psi$.
By Lemma~\ref{lemma-entails}, $J$ entails the candidate invariants
generated at each iteration, thus entails $I$. 
But every model of ${\sf Init}$ (in particular $\A$) is a model of $J$, hence also of $I$. 
Contradiction. Therefore, the assumption that there exists a universal inductive
invariant $J$ that entails $\Psi$
was false, i.e.\ the answer is correct. \QED

\subsection{Examples}
We illustrate the way the algorithm can be applied on several
examples.

\noindent We first illustrate how Algorithm 2 can be applied to
formulae in ${\sf LI}({\mathbb Z})$, ${\sf LI}({\mathbb Q})$ or ${\sf LI}({\mathbb R})$: 

\begin{ex}
We illustrate the way Algorithm 2 can be 
applied to example 12 from \cite{Kapur15}. 
Consider the transition system $T = ( \Sigma_S, {\sf Init}, {\sf
  Update})$, where $\Sigma_S = \Sigma_0 \cup \{ x, y \}$, where
$\Sigma_0$ is the signature of linear integer arithmetic and $x, y$
are 0-ary function symbols,  ${\sf Init} = (x \approx y \wedge x \approx y +2)$ and 
${\sf Update} = (x \leq y + 1 \rightarrow x' \approx x +
2)$. 
Let $\Psi = (y \leq x \wedge x \leq y + 2)$. 
$\Psi$ holds in the initial state. 
To test invariance under transitions we need to check whether 
$$(y \leq x \wedge x \leq y + 2) \wedge x \leq y + 1 \wedge x' \approx x + 2 \wedge y' \approx y 
\wedge (y' > x' \vee x' > y' + 2)$$ 
is satisfiable. The formula is satisfiable, so $\Psi$ is not an
inductive invariant. After eliminating the primed variables we obtain: 

$$\begin{array}{ll}
(y \leq x \wedge x \leq y + 2 \wedge x \leq y + 1 \wedge y > x + 2) &
\vee \\ 
(y \leq x \wedge x \leq y + 2 \wedge x \leq y + 1 \wedge x + 2 > y +
2) & 
\end{array}$$
The first conjunction is unsatisfiable.
The second conjunction can be simplified to  
$(y < x \wedge x \leq y + 1)$.  
If we negate this formula we obtain 
$$\Gamma := (x \leq y \vee x > y + 1)$$
$\Psi \wedge \Gamma = (y \leq x \wedge x \leq y + 2) \wedge (x \leq y
\vee x > y + 1)
\equiv (x \approx y \wedge x \leq y + 2) \vee (y \leq x \wedge y + 1 < x \wedge x \leq
y + 2))$. 
If the constraints are in ${\sf LI}({\mathbb Z})$, the constraint above is 
equivalent to $(x \approx y \vee x \approx y + 2)$ which can be proved
to be an inductive invariant. 
\end{ex}
We now illustrate how Algorithm 2 can be used on an example 
involving arrays and updates. 

% \begin{center}
\begin{figure}[t]
\centering
\begin{minipage}{.5\linewidth}
\begin{verbatim}
  d1 = 3; d2 = a[4]; d3 = 1; 
  while (nondet()) 
   { d1 = a[d1+1]; 
     d2 = a[d2+1] + (1-d3);
     d3 = d3/2     
   }  
\end{verbatim}
\caption{A simple program}
\label{figure-ex}
\end{minipage}
\end{figure}
% \end{center}

\begin{ex}
Consider the program in Fig.~\ref{figure-ex} (a variation of an example 
from \cite{Rybalchenko}). 
The task is to prove that if $a$ is an array with increasingly sorted elements, 
then the formula $\Psi := d_2 \geq a[d_1 + 1]$ is an
invariant of the program.  
$\Psi$ holds in the initial state since 
$d_2 = a[4] = a[3+1]  = a[d_1 + 1] $. 
In order to show that $\Psi$ is an inductive invariant of the while
  loop, we would need to prove that the following formula is unsatisfiable:

$$\begin{array}{l} 
{\sf Sorted}(a) \wedge d_1' \approx a[d_1 + 1] \wedge d_2 ' \approx a[d_2 + 1] + (1- d_3) \wedge d_3' \approx
d_3/2 \wedge 
d_2 \geq a[d_1 + 1] \wedge d_2' < a[d_1' + 1], ~~
\end{array}$$

\noindent where ${\sf Sorted}(a) := \forall i, j ( i \leq j 
\rightarrow a[i] \leq a[j])$. 
The updates change only constants and $\Psi$ is a ground formula. 
If ${\cal T} =  {\mathbb Z} \cup {\sf
  Sorted}(a)$ and  $G = d_2 \geq a[d_1 + 1]
\wedge d_1' \approx a[d_1 + 1] \wedge d_3' \approx d_3/2 \wedge d_2 ' \approx a[d_2 + 1]
+ (1- d_3) $, 
${\cal T} \wedge G$ is satisfiable iff the formula 
$\exists d_1 \exists d_2 \exists d_3 \exists d'_1 \exists d'_2 \exists
d'_3 G$ is valid w.r.t.\ ${\cal T}$. Satisfiability can be checked
using hierarchical reasoning. 

\noindent The (existentially) quantified variables $d_1', d_2'$ and $d_3'$ can
be eliminated already at the beginning. We obtain:  
$$\begin{array}{ll} 
{\sf Sorted}(a) \wedge & d_2 \geq a[d_1 + 1]  \wedge  a[d_2 + 1] + (1- d_3)  < a[a[d_1 + 1] + 1]
\end{array}$$The axiom ${\sf Sorted}(a)$ defines a local theory extension $ {\mathbb Z} \subseteq {\mathbb Z} \cup {\sf
  Sorted}(a) = {\cal T}$; after
flattening of the ground part and instantiation of ${\sf Sorted}(a)$ we obtain: 

\smallskip
$\begin{array}{@{}ll}
G: & c_1 \approx a[d_1 + 1] \wedge d_2 \geq a[d_1 + 1] \wedge a[d_2 + 1] + (1- d_3)  < a[c_1 + 1]\\
{\sf Sorted}(a)[G]: &  d_1 + 1 \rhd d_2 + 1 \rightarrow a[d_1 + 1] \rhd a[d_2 + 1] \\
                           &  d_1 + 1 \rhd c_1 + 1 \rightarrow a[d_1 + 1] \rhd a[c_1 + 1] \\
                           & d_2 + 1 \rhd c_1 + 1 \rightarrow a[d_2 + 1] \rhd a[c_1 + 1], \rhd \in \{ \leq, \geq \} \\
\end{array}$ 

\smallskip
\noindent After purification in which the definitions ${\sf Def}:= \{
c_1 \approx a[d_1 + 1], c_2 \approx a[d_2 + 1], c_3 \approx a[c_1 + 1] \}$ are
introduced and further simplification we obtain: 

\smallskip
$\begin{array}{@{}ll}
G_0: & d_2 \geq c_1\wedge c_2 + (1- d_3)  < c_3 \\
{\sf Sorted}(a)[G]_0: & d_1 \rhd d_2 \rightarrow c_1 \rhd c_2 ~~
\wedge ~~ d_1 \rhd c_1  \rightarrow c_1 \rhd c_3 ~~ \wedge ~~  d_2
\rhd c_1 \rightarrow c_2 \rhd c_3, ~~\rhd \in \{ \leq, \geq \} \\
\end{array}$ 

\

\noindent By negating the formula above and universally
quantifying the constants we obtain a formula $\Gamma$ that 
can be used to strengthen the invariant. The algorithm continues with
the next iteration. Refinements and also 
non-termination are discussed 
later, in Example~\ref{ex-rybal-termination}. 

If we want to find a formula  containing only the variables $d_1,
d_2$ and $d_3$ to strengthen $\Psi$ we can eliminate in addition to
the primed variables also 
$c_1, c_2$ and obtain $d_3 > 1$. 
By negating this condition we obtain $d_3 \leq 1$. 
We obtain 
$d_2 \geq a[d_1 + 1] \wedge d_3 \leq 1$, which can
be proved to be a loop invariant. 
\end{ex}

\

\section{Refinements} 
\label{refinements}

\noindent 
Assume that ${\sf Update} = \bigvee_{f \in F} {\sf Update}_{f}$,
where $F \subseteq \Sigma$ (no $f'$ with $f\in F$ is a parameter) such
that ${\sf Update}_{f}$ satisfies
the conditions in assumption~({\bf A3}).

\begin{lem}
We consider the computations described in Algorithm 2, iteration $n$, in Step 2, 
the case in which ${\sf Init}
\models I_n$, but $I_n$ is not invariant under updates. 
Let $\K$ be a set of constraints on parameters. 
\begin{itemize}
\item[(1)] If $I_n = I_{n-1} \wedge \Gamma_{n-1}$ is not invariant under
updates, then Algorithm~2 
computes a formula $\Gamma_n = \bigwedge_{f \in F} \Gamma^f_n$, where
$\Gamma^f_n$ is obtained by symbol elimination 
applied to $\K \wedge I_n \wedge {\sf Update}_f \wedge G$, where $G$
is obtained by Skolemization from $\neg \Gamma'_{n-1}$. 

\item[(2)] If the only non-parametric functions are $\{ f' \mid f \in F \}$, 
then with the improvement of Algorithm~1  in
Thm.~\ref{symb-elim-simplif} 
we need to apply symbol elimination only 
to ${\sf Update}_f \wedge \neg \Gamma'_{n-1}$  to compute $\Gamma^f_n$. 
\end{itemize}
\label{consider-only-gamma}
\end{lem}
{\em Proof:} 
(1) $I_n \wedge {\sf Update} \wedge \neg I'_n  \equiv  
\bigvee_{f \in F} (I_n \wedge {\sf Update}_f \wedge \neg I'_n)$, 
so it is satisfiable 
iff for some $f \in F$, the formula $\K \wedge I_n \wedge {\sf Update}_f \wedge
\neg I'_n$ is satisfiable. We have: 

\smallskip
\noindent $\begin{array}{lcl} 
\K \wedge I_n \wedge {\sf Update}_f \wedge \neg I'_n & = & \K \wedge (I_{n-1} \wedge \Gamma_{n-1}) \wedge {\sf Update}_f \wedge
(\neg I'_{n-1} \vee \neg \Gamma'_{n-1}) \\
& \equiv & \K \wedge (I_{n-1} \wedge \Gamma_{n-1}) \wedge {\sf Update}_f
\wedge \neg \Gamma'_{n-1}.
\end{array}$ 

\smallskip
\noindent since $\Gamma_{n-1}$ was introduced such that 
$\K \wedge (I_{n-1} \wedge \Gamma_{n-1}) \wedge {\sf Update}_f
\wedge \neg I'_{n-1}$ is unsatisfiable.  Then in Algorithm~2, $\Gamma_{n} = \bigwedge_{f
  \in F} \Gamma^f_n$, where $\Gamma^f_n$ are the (weakest) formulae obtained with 
Algorithm~1, such that $\K \wedge I_n \wedge \Gamma^f_n \wedge {\sf Update}_f \wedge
\neg \Gamma'_{n-1}$ is unsatisfiable. 

\noindent (2) follows from Thm.~\ref{symb-elim-simplif}. \QED

\begin{lem}
If $\phi_i \wedge \phi_j \models_{\cal T} \perp$ for all $i \neq j,
1\leq i, j \leq n$ and
$\models_{\cal T} \bigvee_{i = 1}^n \phi_i$ then 
$\bigwedge_{i = 1}^n (\phi_i\rightarrow C_i) \equiv \bigvee_{i = 1}^n
(\phi_i \wedge C_i)$.  
\label{lemma-DNF}
\end{lem}
\noindent We now analyze the
formulae $\Gamma^f_n$ generated at iteration $n$. 
For simplicity we assume that $f$ is unary; 
the extension to higher
arities is immediate. 
\begin{thm}
Let $\Psi \in {\sf LocSafe}$ and ${\sf Update} = \bigvee_{f \in F}
{\sf Update}_f$ of the form discussed above. Assume that the clauses
in $\Psi$ and ${\sf Update}_f$ are flat and linear for all $f \in F$.  
Let $m$ be the maximal number of variables in a clause in $\Psi$.
Assume that the only non-parametric functions which need to be
eliminated are the primed symbols $\{ f' \mid f \in F \}$ and that 
conditions ({\bf A1})--({\bf A5}) hold. 
Consider a variant of Algorithm~2, which uses for symbol elimination Algorithm~1 with the improvement 
in Theorem~\ref{symb-elim-simplif}.  Then for every step $n$, (i) the clauses in the candidate invariant
$I_n$ obtained at step $n$ of Algorithm~2 are flat, and (ii) the number of universally quantified variables in 
  every clause in $I_n$ is  $\leq m$.
\label{exhaustive-updates}
\end{thm}
{\em Proof:} Proof by induction on $n$. For $n = 1$, $I_1 =
\Psi$ and  (i) and (ii) clearly hold. Assume that they hold  
for iteration $n$. We prove that they hold for iteration $n+1$. 
By Lemma~\ref{consider-only-gamma},  
we need to apply Algorithm~1
to ${\sf Update}_f \wedge G$, where
$G$ 
is obtained from $\neg \Gamma'_n$ after Skolemization. If $\Gamma'_n$ 
is a conjunction of clauses, then $G$ is a disjunction of
conjunctions of literals; each disjunct can be processed separately,
and we take the conjunction of the obtained constraints. Thus, we
assume w.l.o.g.\ that $G$ is a conjunction of literals. 
By the induction hypothesis the number $k$ of universally quantified
variables in $\Gamma_n$ is $\leq m$, so $G$ contains 
Skolem constants $\{ d_1, \dots, d_k, c_1, \dots, c_r \}$ with $k+r
\leq m$,  where $d_1, \dots, d_k$ occur below $f'$.
For symbol elimination we first compute 
$G_1 = {\sf Update}_f[G] \wedge G$ 
(with $ {\sf est}(G) = \{
f'(d_1), \dots, f'(d_k) \}$ where $k \leq m$) and purify it; 
in a second step we instantiate the terms starting with function symbols 
$g \in \Sigma_P \cup \Sigma$.  
By Lemma~\ref{lemma-DNF}: 

\smallskip
\noindent $\begin{array}{@{}r@{}c@{}l} 
{\sf Update}_f[G] & := & 
\!\!\displaystyle{\bigwedge_{j = 1}^{k} \!\left( \!\bigwedge_{i = 1}^{n_f}
  (\phi_i(d_j) \rightarrow C_i(d_j, f'(d_j))) \!\!\right) 
\equiv
\bigwedge_{j = 1}^{k} \bigvee_{i = 1}^{n_f} (\phi_i(d_j) \wedge C_i(d_j,
f'(d_j))) } \\
& \equiv & \displaystyle{\bigvee_{i_1, \dots, i_k \in \{ 1, \dots,
    n_f\} }
\left(\bigwedge_{p = 1}^k \phi_{i_p}(d_p)  \wedge
\bigwedge_{p = 1}^k C_{i_p}(d_p, f'(d_p)) \right).} 
\end{array}$

\smallskip
\noindent We thus obtained a DNF with $(n_f)^k \leq (n_f)^m$ disjuncts, where $n_f$
(number of cases in the definition of $f$) and
$m$ (the maximal number of variables in $\K \cup I_n$) 
are constants depending on the description of the transition system. 
Both $n_f$ and $m$ are typically small, in most cases 
$n_f \leq 3$. Algorithm~1 is applied as follows: 

\medskip
\noindent In {\bf Step 1}   
we introduce a constant $c_{f'd}$ for every term $f'(d) \in {\sf
  est}(G)$, replace  $f'(d)$ with $c_{f'd}$, and add the
corresponding instances ${\sf Con}_0$ of the congruence axioms. 
We may compute a disjunctive normal form $DNF({\sf Con}_0)$ for the
instances of congruence axioms or not (${\sf Con}_0$ 
contains $k^2 \leq m^2$
conjunctions;  
$DNF({\sf Con}_0)$ contains $2^{k^2} \leq 2^{m^2}$ disjuncts,
each of length $k$). 
In a second reduction we replace every term of the form
$g(c) \in {\sf est}(G_1)$, $g \in \Sigma_P$, with a new constant
$c_{gc}$. 

\medskip
\noindent {\bf Steps 2 and 3:} To eliminate $f'$ 
we replace the constants $c_{f'd}$ with variables $x_{f'd}$
and obtain a formula $G^0_1(x_{f'd_1}, \dots, x_{f'd_n})$. 
In $\exists x_{f'd_1}, \dots x_{f'd_n} ~ G^0_1(x_{f'd_1}, \dots, x_{f'd_n})$ 
the existential quantifiers can be brought inside the conjunctions and quantifier elimination can
be used only on the part of the disjuncts that contain the variables
$x_{f'd}$ (i.e.\ on relatively simple and short formulae). 
After quantifier elimination we obtain a formula $\Gamma_2$.

\medskip
\noindent {\bf Steps 4 and 5:} We replace back in the formula obtained this
way all constants $c_{gc}$, $g \in \Sigma_P, g(c) \in {\sf est}(G_1)$
with the terms $g(c)$.  The constants $d_1, \dots, d_k, c_1, \dots, c_r$ are replaced
with new variables $y_1, \dots, y_k,$ $y_{k+1}, \dots, y_{k+r}$ respectively. 
We negate $\exists {\overline y} 
\Gamma_2({\overline y})$ and obtain a conjunction $\Gamma^f_{n+1}(G)$ 
of universally quantified clauses.

\medskip
\noindent All clauses in $\Gamma^f_{n+1}(G)$ are flat. 
By construction, the number of universally quantified 
variables in $\Gamma^f_{n+1}(G)$ is $k + r \leq m$. 
A full proof is included in Appendix~\ref{app:exhaustive-updates}. 
\QED

\begin{thm}
Under the assumptions in Theorem~\ref{exhaustive-updates},
the number of clauses in $\Gamma_n$ is at most $O(k_1^n)$; 
each clause in $\Gamma_n$ contains at most
$k_2 \cdot n + |\Psi|$ literals if the constraints $C_i$ are all equalities,
and can contain $O(|\Psi|^{k_3^n})$ literals 
if $C_i$ are constraints in ${\sf LI}({\mathbb Q})$, 
where $k_1, k_2, k_3$ are constants of the system. 
\label{thm:length-formulae}
\end{thm}

\noindent {\em Proof:} 
We first analyze the number $c(n)$ of clauses in $\Gamma_n$. 
$\Gamma_n = \bigwedge_{f \in F} \Gamma^f_n$.  
From the proof of Theorem~\ref{exhaustive-updates} it can be seen 
that for every clause $C$ in $\Gamma_n$, the number of clauses in 
$\Gamma^f_{n+1}$ generated to ensure unsatisfiability of 
${\sf Update}_f \wedge G$, where $G = \neg C$ is at most $n_f^m * m^2
\leq {\sf mc}^m * m^2$, where ${\sf mc}$ is the maximal number of cases used 
for the updates of the functions in $f$ (i.e.\ $n_f \leq {\sf mc}$ for all
$f \in F$).  
Thus, the number $c(n+1)$ of clauses in $\Gamma_{n+1}$ 
satisfies $c(n+1) \leq |F| * {\sf mc}^m * m^2 * c(n) = k_1 * c(n)$. This shows that 
$c(n+1) \leq k_1^n*c(1)$, where $k_1 = |F|* {\sf mc}^m * m^2$ is a constant of
the system.

We now analyze the length of the clauses in $\Gamma_n$. 
Let $p$ be the maximal length\footnote{i.e. the
  maximal number of literals}
of the constraints $C_i$, and $p(n)$
the maximal length of the clauses in $\Gamma_n$. Then the negation of 
every clause in $\Gamma_n$ contains at most $p(n)$ literals; after
instantiation and DNF transformation we obtain a disjunction of 
conjunctions of the form 
$$\bigwedge_{i = 1}^k \phi_{i_p}(d_p) \wedge \bigwedge_{i = 1}^k
C_{i_p}(d_p, c_{f'd_p})) \wedge \bigwedge_{(f'(d_i),f'(d_j)) \in D} d_i \not\approx d_j \wedge
  \bigwedge_{(f'(d_i),f'(d_j)) \in CD} c_{f'd_i} \approx c_{f'd_j} \wedge G_0$$
(where $D \subseteq {\sf est}(G)^2$ and $CD = {\sf est}(G)^2
\backslash D$) with at most $(k * l_f + k^2 + |G|) \leq (m *l_f + m^2 +
p(n))$ literals, where $m$ is the maximal number of variables in
$\Psi$ and $l_f$ the maximal length of a clause in ${\sf Update}_f$.
We perform QE essentially on $\bigwedge_{p = 1}^k C_{i_p}(d_p, x_{fd_p}) 
\wedge G$ (the conjunction of equalities 
$x_{f'd_1} \approx x_{f'd_2}$ is processed fast). This conjunction
contains at most $k p + p(n) \leq mp + p(n)$ literals. 

If the constraints $C_i$ are
equalities (as it is the case with updates of the form ${\sf D}_f$) 
the formula obtained by quantifier elimination has equal length or is
shorter than $\bigwedge_{p = 1}^k C_{i_p}(d_p, x_{fd_p})
\wedge G$, so $p(n+1) \leq m*p + p(n)$, i.e.\ $p(n) \leq k_2 n +
|\Psi|$, where $k_2 = m*p$.  

If the constraints
are conjunctions of linear inequalities over ${\mathbb Q}$ then using 
e.g.\ the Fourier-Motzkin elimination procedure, after eliminating one 
variable, the size is at most $(m*p + p(n))^2$, 
after eliminating $m$ variables it is $(m*p + p(n))^{2^m}$. 
Thus, $p(n+1) \leq (k_2 + p(n))^{k_3}$, where 
$k_3 = 2^m$
is a constant depending on the system. 
Thus, $p(n)$ is in the worst case in $O(p(1)^{k_3^n})$.

\smallskip
\noindent However, in many of the examples we analyzed the growth of
formulae is not so dramatic because (i) in many cases the updates are
assignments, (ii) even if the updates are specified by giving lower
and upper bounds for the new values, the length of the constraints $C_i$ is 1 or 2; 
(iii) many of the disjuncts in the DNF to which quantifier elimination should be 
applied are unsatisfiable and do not need
further consideration. \QED

\smallskip 
\noindent If there are non-parametric functions that are being updated the
number of variables in the clauses $\Gamma_n$ might grow: Any constant
$c \in F$ which is not a parameter, but occurs below a parameter in
${\sf Update}$ or $G$, is then being converted into a universally quantified
variable by Algorithm~1 as the following example shows. 
\begin{ex}
Consider the program in the introduction 
(Fig.~\ref{fig2}). 
The task is to prove that if the parameter $b$ is an increasingly sorted array then $\Psi := d_2 \geq d_1$ is an
invariant of the program. $\K_P$ contains the sortedness axiom for
$b$.\\
Assume first that $\Sigma_P = \{ b, d_1,
d_2, a \}$.  $\Psi$ clearly holds in the
initial state. 
To show that $\Psi$ is an inductive invariant of the while
loop, we would need to prove that the following formula is unsatisfiable: 

\smallskip
$
d_1 \leq d_2 \wedge \forall j (a'[j] \approx a[j]+1) \wedge  d'_1
\approx a'[i] \wedge d'_2 \approx a'[i+1] \wedge i' \approx i + 1
\wedge d'_1 >  d'_2. 
$ 

\smallskip
\noindent We have the chain of local theory extensions 

\smallskip
\quad \quad $ {\mathbb Z} ~~\subseteq~~ {\mathbb Z} \cup {\sf UIF}_{a} ~~\subseteq~~ {\mathbb
  Z} \cup {\sf UIF}_{a} \cup {\sf Update}_a = {\cal T},$

\smallskip
\noindent where ${\sf Update}_a = \forall j (a'[j] \approx
a[j]+1)$. ${\sf Update}_{d_1} = d'_1
\approx a'[i]$, ${\sf Update}_{d_2} = d'_2 \approx a'[i+1]$ and ${\sf
  Update}_i = i' \approx i + 1$ are
ground formulae. Let $G =  d'_1
\approx a'[i] \wedge d'_2 \approx a'[i+1] \wedge i' \approx i + 1
\wedge d'_1 >  d'_2$. 
Using the hierarchical reduction method for local theory extensions we
can see that the formula above is satisfiable, so $\Psi$ is not an
invariant. 
To strengthen $\Psi$ we use  Algorithm~1; 
by  Theorem~\ref{symb-elim-simplif}
we can ignore $\K_P$ and $I_1 = d_1 \leq d_2$. 
In a first step, we compute ${\sf Update}_a[G]$ and 
obtain the set of instances $a'[i] \approx a[i] + 1 \wedge a'[i+1]
\approx a[i+1] + 1$. After purification 
we obtain (with ${\sf Def} = a'_1 \approx a'[i] \wedge
a'_2 \approx a'[i+1]$): 
 
\smallskip
\noindent $\begin{array}{l@{}l} 
G_0 \wedge ({\sf Update}_a)_0{:}~ & a'_1 {\approx} a[i] {+} 1 \wedge  a'_2 {\approx} a[i{+}1] {+} 1 \wedge 
d'_1 {\approx} a'_1 \wedge d'_2 {\approx} a'_2 \wedge i' {\approx} i {+} 1
\wedge d'_1 {>} d'_2
\end{array}$

\smallskip
\noindent
In a second step we can use a similar hierarchical reduction for the
extension with ${\sf UIF}_a$; we obtain (with 
${\sf Def} = a'_1 \approx a'[i] \wedge a'_2 \approx a'[i+1] \wedge  a_1 \approx a[i]
\wedge a_2 = a[i+1]$): 

\smallskip
$a'_1 \approx
a_1 + 1 \wedge a'_2 \approx a_2 + 1 \wedge 
 d'_1 \approx a'_1 \wedge d'_2 \approx a'_2 \wedge i' \approx i + 1
\wedge d'_1 > d'_2$.

\smallskip
\noindent
We use quantifier elimination for eliminating $a'_1, a'_2, d'_1, d'_2,$
$i'$ and obtain the constraint $a_1 > a_2$. 
After replacing the constants with the terms they denote 
we obtain $\exists i (a[i] > a[i+1])$; its negation, 
$\Gamma_1 = \forall i (a[i] \leq a[i+1])$, can be used to
strengthen $\Psi$ to the inductive invariant 
$I_2 := d_1 \leq d_2 \wedge \forall i (a[i] \leq a[i+1])$. 
Note that $\Gamma_1$ and hence also $I_2$ contain one universally
quantified variable 
more than $I_1 = \Psi$.

\smallskip
\noindent Assume now that $\Sigma_P = \{ b, d_1, d_2, a, i \}$. All primed
variables are eliminated as above, but in Step 2 of Algorithm~1
$i$ is not existentially quantified. $\Psi$ is strengthened to 
$I_2 := d_1 \leq d_2 \wedge a[i] \leq a[i+1]$ (no universally
quantified variables, the same as in $\Psi$). However, $I_2$ is not an
inductive invariant. It can be strengthened to $I_3 :=  d_1 \leq d_2 \wedge
a[i] \leq a[i+1] \wedge a[i+1] \leq a[i+2]$ and so on. Ideas similar
to those used in the melting
calculus \cite{horbach} (used e.g. in
\cite{horbach-sofronie,horbach-sofronie-2014}) 
could be used to obtain $d_1 \leq d_2 \wedge \forall i (a[i] \leq a[i+1])$. (This example 
indicates that it could be a good strategy to not include the
variables controlling loops among the parameters.) 
\end{ex}
\begin{cor}
The symbol elimination method in Algorithm~1 can be adapted to eliminate all
constants not guarded by a function in $\Sigma_P$. 
With this change we can guarantee that 
 in all clauses in $I_n$ all variables occur below a
  function in $\Sigma_P$. 
\end{cor}
\begin{ex}
Let $\T_0 = LI({\mathbb Q})$. Let $m, M, g, L \in \Sigma_P$ satisfying
$\K = \{ m \leq M \}$. Assume that ${\sf Update} = {\sf Update}_f$, 
where
${\sf Update}_f := \{ \forall x( x \leq c_1 \rightarrow m \leq
f'(x) \wedge f'(x) \leq M), \forall x (x > c_1 \rightarrow f'(x)
\approx a) \}$. Consider the formula $\Psi = \forall x, y (g(y) \leq x \rightarrow
f(x) \leq L(y))$. By the results in Examples~\ref{ex-monotone}
and~\ref{examples-local} we have the following 
chain of local theory extensions: 

\smallskip
$ \quad \quad \quad \T_0 ~~\subseteq~~ \T_0 \cup {\sf UIF}_{\{ g, L\}} ~~\subseteq~~ \T_0 \cup \Psi ~~\subseteq~~ \T_0 \cup \Psi \cup {\sf
  Update}_f.$

\smallskip
\noindent $\Psi$ is invariant under the update of $f$ iff 
$\K \wedge \Psi \wedge {\sf Update}_f \wedge G$  is
unsatisfiable, where $G = g(c) \leq d \wedge f'(d) > L(c)$ is obtained
from $\neg \Psi'$ after Skolemization. 
Since this formula is satisfiable, $\Psi$ is not
invariant. We can strengthen $\Psi$ as explained before, by 
computing the DNF of ${\sf Update}_f[G] = \{ (d \leq c_1 \rightarrow m \leq
f'(d) \wedge f'(d) \leq M), (d > c_1 \rightarrow f'(d) \approx a) \}$,
 as explained in Lemma~\ref{lemma-DNF}, replacing $f'(d)$ with an existentially quantified
variable $x_{f'd}$. Also the constant $d$ does not occur below a
parameter, so it is replaced in Step 2 of Algorithm~1 with a variable
$x_d$. The terms $g(c)$ and $L(c)$ are replaced with constants
$c_{gc}$ and $c_{Lc}$.  We obtain: 

\smallskip
$ \quad \quad \quad \begin{array}{lll} 
\exists x_d \exists x_{f'd} & (x_d \leq c_1 \wedge m \leq x_{f'd} \wedge x_{f'd}
\leq M \wedge c_{gc} \leq x_d \wedge x_{f'd} > c_{Lc}) & \vee \\
&  (x_d > c_1 \wedge
x_{f'd} \approx a \wedge c_{gc} \leq x_d \wedge x_{f'd} > c_{Lc})
\end{array}$

\smallskip
\noindent After eliminating $x_{f'd}$ we obtain: 

\smallskip
\noindent $\exists x_d  (x_d \leq c_1 \wedge c_{gc} \leq x_d \wedge m \leq M \wedge c_{Lc} < M)
\vee (x_d > c_1 \wedge c_{gc} \leq x_d \wedge c_{Lc} < a).$

\smallskip
\noindent The variable $x_d$ does not occur below any function symbol and it can be
eliminated; we obtain the equivalent formula 
$(c_{gc} {\leq} c_1 \wedge m {\leq} M \wedge c_{Lc} {<} M) \vee (c_{Lc} {<} a)$; 
after replacing back the constants $c_{gc}$ and $c_{Lc}$ with the
terms they denote and replacing $c$ with a new
existentially quantified variable $y$ (Step 4 of Algorithm~1) and negating the formula obtained
this way (Step 5 of Algorithm~1) we obtain the constraint $\forall y (g(y) \leq c_1
\rightarrow M \leq L(y)) \wedge \forall y (a \leq L(y))$ which we can
use to strengthen $\Psi$ to an inductive invariant.
\end{ex}

\subsection{Avoiding some of the conditions ({\bf A1})--({\bf A5})}
\label{no-a5}

Assumption ${\bf (A4)}$ ($\T_0$ allows quantifier elimination) 
is not needed 
if in all update axioms 
 $f'$ is defined using equality; then $f'$
can easily be eliminated. 

\smallskip
\noindent 
Assumption {\bf (A5)} is very strong. 
Even if we cannot guarantee that assumption {\bf  (A5)} holds, it could
theoretically be possible to identify situations in which  we can
transform candidate invariants which do not define local extensions
into equivalent formulae
which define local extensions -- e.g.\ using 
the results in \cite{horbach-sofronie}. 

\smallskip
\noindent If all candidate invariants $I$ generated
in Algorithm~2 are ground, assumption  {\bf (A5)} is not
needed. 

\medskip

\noindent In what follows we describe a situation in which assumption {\bf
  (A5)} 
is fulfilled, 
so Lemma~\ref{lemma-entails} and Theorem~\ref{inv-gen-correctness} hold under assumptions {\bf
  (A1)}--{\bf (A4)}.

\medskip
\noindent 
We consider transition systems $T = (\Sigma_S, {\sf Init}, {\sf
  Update})$ and properties $\Psi$, 
where $\T_S = \T_0 \cup \K$ and 
${\sf Init}, \Psi, \K$ and ${\sf Update}_f$, for $f \in F$, are all in
the {\em array property fragment} (APF). 
Then assumptions $({\bf A1})$ and $({\bf A2})$ hold. 
We identify conditions under which we can guarantee that at every iteration 
of Algorithm~2, the candidate invariant $I$ is in 
the array property fragment, so the locality assumption {\bf
  (A5)} is fulfilled and does not need to be mentioned explicitly. 

\smallskip
\noindent Many types of systems have descriptions in this fragment; an
example follows. 
\begin{ex}
\label{example-water-tank}
Consider a controller of a water tank in which 
the inflow and outflow in a time unit can be chosen freely between
minimum and maximum values that depend 
on the moment in time. 
At the beginning minimal and maximal values for the inflow and outflow are
initialized as described by the formula ${\sf Init} = {\sf In_1}
\wedge {\sf In_2} \wedge {\sf Out}_1 \wedge {\sf Out}_2 \wedge (t
\approx 0) \wedge (L \approx L_0)$, where for $i = 1, 2$: 

\smallskip

$~{\sf In_i}  = \forall t (0 \leq {\sf in}^i_m(t) \leq  {\sf in}^i_M(t) \leq 
L_{\sf overflow} {-} L_{\sf  alarm} {-} \epsilon_i)$ and 

${\sf Out_i}  = \forall t ({\sf in}^i_M(t) \leq {\sf
    out}^i_m(t) \leq {\sf out}^i_M(t))$

\smallskip
\noindent The updates are described by ${\sf
  Update}_{\sf in} \wedge L' = L {+} {\sf in'}(t) \wedge t' = t {+} 1$, where: 

\smallskip
\noindent $\begin{array}{rcl}
{\sf Update}_{\sf in} & = & \forall t (L \leq L_{\sf alarm} \wedge t \leq
t_0 \rightarrow {\sf in}^1_m(t) \leq {\sf in}'(t) \leq {\sf in}^1_M(t)) \\
& & \forall t (L \leq L_{\sf alarm} \wedge t >
t_0 \rightarrow {\sf in}^2_m(t) \leq {\sf in}'(t) \leq {\sf in}^2_M(t)) \\
& & \forall t (L > L_{\sf alarm} \wedge t \leq
t_0 \rightarrow {\sf in}^1_m(t) {-} {\sf out}_M^1(t) \leq {\sf in}'(t) \leq
{\sf in}^1_M(t) {-} {\sf out}^1_m(t)) \\
& & \forall t (L > L_{\sf alarm} \wedge t >
t_0 \rightarrow {\sf in}^2_m(t) {-} {\sf out}^2_M(t) \leq {\sf in}'(t) \leq
{\sf in}^2_M(t) {-} {\sf out}^2_m(t)) 
\end{array}$

\smallskip
\noindent Assume that $\Sigma_P = \{ L_{\sf alarm}, L_{\sf overflow},
L, L_0, {\sf in}^1_m, {\sf in}^1_M, {\sf in}^2_m, {\sf in}^2_M, 
{\sf out}^1_m, {\sf out}^1_M, {\sf out}^2_m, {\sf out}^2_M \}$, 
and let 
$\K_P = \{ 0 \leq L_{\sf alarm}, L_{\sf alarm} < L_{\sf overflow},
L_0\leq L_{\sf alarm} \}$. 
All these formulae are in the array property fragment. 
In fact, these formulae satisfy also the conditions in
Example~\ref{examples-local}, so we have the following local theory
extensions: 
$$ \T_0 \cup \K_P ~~\subseteq~~ \T_0 \cup \K_P \cup {\sf Init} \quad \quad
\text{ and } \quad \quad \T_0 \cup \K_P ~~\subseteq~~
\T_0 \cup \K_P \cup {\sf Update}_{\sf in}.$$

\noindent 
We illustrate the way the refinement of Algorithm~2 described in
Section~\ref{refinements} can be used to strengthen the
property $\Psi = L \leq L_{\sf overflow}$. 

\

\noindent 
Let $I_1 := \Psi$.  It is easy to check that ${\sf Init} \models
I_1$. Indeed, $\K_P \wedge {\sf Init} \wedge L > L_{\sf overflow}
\models \perp$ because already $L_0 \leq L_{\sf alarm} \wedge L_{\sf
  alarm} < L_{\sf overflow} \wedge L \approx L_0 \wedge L > L_{\sf
  overflow}$ is unsatisfiable. 
We now check whether $I_1$ is invariant under updates, i.e.\ 
whether $I_1 \wedge {\sf Update} \wedge \neg I_1$ is 
satisfiable or not.  We can show that 
$\K_P \wedge L \leq L_{\sf overflow} ~\wedge~  {\sf
  Update}_{\sf in} \wedge L' = L {+} {\sf in'}(t) \wedge t' = t {+} 1
\wedge L' > L_{\sf overflow}$ is satisfiable using  
hierarchical reasoning in the chain of local theory extensions
mentioned above, 
so $I_1$ is not an invariant. We strengthen $I_1$ by applying 
the improvement of the symbol elimination algorithm, according to
Theorem~\ref{symb-elim-simplif},  
to 
$${\sf
  Update}_{\sf in} \wedge L' = L {+} {\sf in'}(t) \wedge t' = t {+} 1
\wedge L' > L_{\sf overflow}$$
After instantiating the universally quantified variables $j$ with $t$,
using  Lemma~\ref{lemma-DNF} we can bring ${\sf Update}_{\sf in}$ to
DNF, using distributivity, and replacing the ground terms $f(t)$ with
$c_{ft}$ and ground terms to be eliminated ${\sf in}'(t), L', t'$ with
variables $x_{in}, L', t'$ we obtain the following disjunction 
of formulae: 

\smallskip
\noindent $\begin{array}{@{}l@{}l@{}l}
& \displaystyle{\bigvee_{i = 1}^2} & (L \leq L_{\sf alarm} \wedge C_i(t) \wedge
c_{{\sf in}^i_m(t)} \leq x_{in}  \leq c_{{\sf in}^i_M(t)} \wedge
L' \approx L + x_{in} \wedge t' \approx t + 1 \wedge L' > L_{\sf overflow}) \\
\vee & \displaystyle{\bigvee_{i = 1}^2} & (L > L_{\sf alarm} \wedge C_i(t) \wedge
c_{{\sf in}^i_m(t)} - c_{{\sf out}^i_M(t)} \leq x_{in} \leq c_{{\sf
    in}^i_M(t)} - c_{{\sf out}^i_m(t)}\wedge
L' \approx L + x_{in} \wedge t' \approx t + 1 \\
& & ~\wedge L' > L_{\sf overflow})
\end{array}$

\smallskip
\noindent (for uniformity we use the notation  $C_1(t) := t \leq t_0, C_2(t) :=
t > t_0$). \\
After eliminating the existentially quantified variables
$x_{in}, L', t'$ and replacing $c_{gc}$ back with $g(c)$ we obtain:

\smallskip
\noindent $\begin{array}{@{}l@{}l@{}l}
& \displaystyle{\bigvee_{i = 1}^2} & (L \leq L_{\sf alarm} \wedge
C_i(t) \wedge
{\sf in}^i_m(t) \leq {\sf in}^i_M(t) \wedge L_{\sf overflow} - L < {\sf in}^i_M(t)) \\
\vee & \displaystyle{\bigvee_{i = 1}^2} & (L > L_{\sf alarm} \wedge C_i(t) \wedge
{\sf in}^i_m(t) {-} {\sf out}^i_M(t) \leq {\sf in}^i_M(t) {-} {\sf
  out}^i_m(t) \wedge L_{\sf overflow} {-} L  < {\sf in}^i_M(t) {-} {\sf
  out}^i_m(t) ). 
\end{array}$

\smallskip 
\noindent 
$t$ is not a parameter, so we consider it to be existentially
quantified. 
After negation we obtain the formula $\Gamma_1$ (equivalent
to a formula in the
array property fragment): 

\smallskip
\noindent 
$\begin{array}{@{}l@{}l}
  \forall t & (L {\leq} L_{\sf alarm} \wedge
t \leq t_0 \wedge
{\sf in}^1_m(t) {\leq} {\sf in}^1_M(t) \rightarrow  {\sf in}^1_M(t)
\leq L_{\sf overflow} - L) \\
 \forall t & (L {\leq} L_{\sf alarm} \wedge
t > t_0 \wedge
{\sf in}^2_m(t) {\leq} {\sf in}^2_M(t) \rightarrow  {\sf in}^2_M(t)
\leq L_{\sf overflow} - L) \\
\forall t & (L {>} L_{\sf alarm} \wedge t \leq t_0 \wedge
{\sf in}^1_m(t) {-} {\sf out}^1_M(t) {\leq} {\sf in}^1_M(t) {-} {\sf
  out}^1_m(t) \rightarrow {\sf in}^1_M(t) {-} {\sf
  out}^1_m(t)  \rightarrow L_{\sf overflow} {-} L)  \\
\forall t & (L {>} L_{\sf alarm} \wedge t > t_0 \wedge
{\sf in}^2_m(t) {-} {\sf out}^2_M(t) {\leq} {\sf in}^2_M(t) {-} {\sf
  out}^2_m(t) \rightarrow {\sf in}^2_M(t) {-} {\sf
  out}^2_m(t)  \rightarrow L_{\sf overflow} {-} L). 
\end{array}$

\smallskip
\noindent It can be checked again using hierarchical reasoning in local theory
extensions that ${\sf Init} \models \Gamma_1$, so $I_2 = I_1 \wedge
\Gamma_1$ holds in the initial states. To prove that $I_2$ is an
inductive invariant note that 

\smallskip
$\begin{array}{lclcl}
& (I_2 & \wedge {\sf Update}_{\sf in} \wedge L' \approx L + {\sf in}(t)
\wedge t' \approx t + 1 \wedge & \neg I_2') &  \\
\equiv & ((I_1 \wedge \Gamma_1) & \wedge {\sf Update}_{\sf in} \wedge L' \approx L + {\sf in}(t)
\wedge t' \approx t + 1 \wedge & (\neg I_1' \vee \neg \Gamma'_1)) & \\
\equiv & ((I_1 \wedge \Gamma_1) & \wedge {\sf Update}_{\sf in} \wedge L' \approx L + {\sf in}(t)
\wedge t' \approx t + 1 \wedge & \neg I_1') & \vee  \\
&  ((I_1 \wedge \Gamma_1) & \wedge {\sf Update}_{\sf in} \wedge L' \approx L + {\sf in}(t)
\wedge t' \approx t + 1 \wedge & \neg \Gamma_1) & \\
\equiv & \perp. 
\end{array}$ 

\smallskip
\noindent because (1) by the construction of $\Gamma_1$, $((I_1 \wedge
\Gamma_1) 
\wedge {\sf Update}_{\sf in} \wedge L' \approx L + {\sf in}(t)
\wedge t' \approx t + 1 \wedge  \neg I_1')$ is unsatisfiable, and (2)
$\Gamma_1$ does not contain any function which is updated, so
$\Gamma'_1 = \Gamma_1$. 
 \end{ex}
The following results follow from the definition of the array property
fragment.
\begin{lem}
Under assumption $({\bf
  A3})$, ${\sf Update}_f$ is in the array property fragment iff 
$\phi_1, \dots \phi_{n_f}$ are conjunctions of constraints of the form
$x \leq g$ or $x \geq g$, where $x$ is a 
variable and $g$ is a ground term of sort
${\sf index}$, all $\Sigma \cup \Sigma_P$ terms are flat and 
all universally quantified variables occur below a function in $\Sigma
\cup \Sigma_P$. 
\end{lem}
{\em Proof:} Follows from the definition of the array property
fragment. \QED
\begin{lem}
Let $G$ be the negation of a formula in the array property fragment (APF). 
Then the following are equivalent: 
\begin{itemize}
\item[(1)] The formula obtained by applying Algorithm~1 to ${\sf Update}_f \wedge
G$ is in the APF.  
\item[(2)] No instances of the
congruence axioms need to be used for ${\sf
  est}(G)$. 
\item[(3)] Either ${\sf est}(G)$ contains only one element,
or whenever $f'({\overline d}), f'(\overline{d'}) \in {\sf
  est}(G)$, where ${\overline d} = d_1, \dots, d_n$, $\overline{d'} =
d'_1, \dots, d'_n$, we have $\T_0 \cup \K \cup G \models \bigvee_{i = 1}^n d_i \not\approx d'_i$. 
\end{itemize}
\end{lem}
{\em Proof:} Consequence of the fact that the formula
$\displaystyle{\!\! \bigwedge_{p = 1}^k \phi_{i_p}(y_p)  \wedge \!\!\!\!\!\!\!\!\bigwedge_{(d_1, d_2) \in D} 
\!\!\!\!\!\!y_1 \not\approx y_2 \wedge G^g_0({\overline y}, {\overline
  g(y)})}$ obtained after
applying Algorithm~1,   can be an index guard only if it does not contain the
disequalities $y_1 \not\approx y_2$. 
This is the case when $|{\sf est}(G)| = 1$ or else if for all $f(d_1), f(d_2) \in {\sf est}(G)$, $\T_0 \cup
G_0 \models d_1 \not\approx d_2$. \QED

\begin{thm}
Let $T = (\Sigma_S, {\sf Init}, {\sf Update})$ be a transition system
with theory $\T_S = \T_0 \cup \K$. Assume that $\T_0$ is the disjoint
combination of Presburger arithmetic (sort ${\sf index}$) and a theory 
of elements (e.g.\ linear arithmetic over ${\mathbb
  Q}$). Assume that all functions in $\Sigma$ are unary. 
If $\K, {\sf Init}, {\sf Update}$ and $\Psi$ are in the array
  property fragment and all clauses in $\Psi$ have only one
  universally quantified
  variable, 
then the formulae $\Gamma_n$ obtained by symbol elimination in 
Step 2 at every iteration of Algorithm~2 are again in the array property
fragment and are conjunctions of clauses having only one quantified
variable. 
\end{thm}

\subsection{Termination}
\label{termination}

Algorithms of the form of Algorithm~2 do not terminate in general even for
simple programs, handling only integer or rational variables (cf.\
e.g.\ \cite{Dillig}). 
We identify situations in which the invariant synthesis procedure terminates.

\begin{lem}[A termination condition]
Assume that the candidate invariants $I$ generated at each iteration are
conjunctions of clauses which contain, up to renaming of the
variables, terms in a
given finite family ${\sf Ter}$ of terms. Then the
algorithm must terminate with an invariant $I$ or after detecting that
${\sf Init} \not\models I$. 
\label{termination-general}
\end{lem}

\noindent {\em Proof:} Assume that this is
not the case. Then -- as only finitely many clauses can be generated -- at some iteration  
$n$ a candidate invariant $I_n = I_{n-1} \wedge
\Gamma_{n-1}$ equivalent to $I_{n-1}$ is obtained. Then, on the one 
hand $I_{n-1} \wedge {\sf Update} \wedge \neg
I'_{n-1}$ is satisfiable, on
the other hand, being equivalent to $\Gamma_{n-1} \wedge I_{n-1} \wedge
{\sf Update} \wedge \neg I'_{n-1}$ it must be
unsatisfiable. Contradiction. \QED

\medskip
\noindent A situation in which this condition holds is described below.\footnote{To simplify the notation, we assume
that the functions in $\Sigma$ have arity $\leq 1$. Similar arguments
can be used  
for $n$-ary functions. } 

\begin{thm}
Let $\Sigma = \{ f_1, \dots, f_n \}$. Assume that $\Sigma_P = \Sigma$,
$\T_0 = {\sf LI}({\mathbb Q})$ and 
that: 
\begin{itemize}
\item All clauses used for defining ${\cal T}_S$ and 
the property $\Psi$ contain only literals of the form: 
$x \rhd t$, $u \rhd v$, $f_i(x) \rhd s$, $f_i(x) \rhd y$, where 
$x, y$ are (universally quantified) variables, 
$f_i \in \Sigma$, $s, t, u, v$ are constants, and $\rhd \in \{ \leq, <, \geq, >, \approx \}.$ 
\item All axioms in ${\sf Update}$ are of the form 
$ \forall {\overline x} \left( \phi^k_i({\overline x}) \rightarrow
 C_i({\overline x}, f'_k({\overline x})) \right)$ as in
Assumption~({\bf A2}), 
where  $C_i({\overline x}, y)$ 
and $\phi^k_i({\overline x})$ are conjunctions of literals of
the form above. % , and $s^k_i, t^k_i$ are either variables or constants. 
\end{itemize}
Then all the candidate
invariants $I$ generated during the execution of the algorithm in 
Fig.~\ref{fig-inv-gen} are equivalent to sets of clauses, all containing
a finite
set ${\sf Ter}$ of terms formed with variables in a finite set ${\sf Var}$. 
Since only finitely many clauses (up to renaming of variables) can be
formed this way, after a finite number of steps no 
new formulae can be generated, thus the algorithm terminates. 
\label{examples-termination}
\end{thm}
\noindent {\em Proof:} 
We analyze the way the formulae $\Gamma$ are built using Algorithm 2. 
We prove by induction that for every $n \in {\mathbb N}$ the candidate 
invariant $I_n$ constructed in the $n$-th iteration 
is a conjunction of clauses, 
each clause having universally quantified variables in the finite set
${\sf Var}$ and terms obtained from the terms in ${\sf Ter}$ by
possibly renaming the variables. 

\smallskip
\noindent 
In the first iteration, $I_1 = \Psi$, so the property holds. 
Assume that the property holds for $I_n$. 
Consider the case in which  
${\sf Init} \models I_n$ and $I_n$ is not invariant under all updates,
i.e.\ there exists $f \in \Sigma$ such that 
$I_n \wedge {\sf Update} \wedge G$ is satisfiable, 
where $G$ is obtained from $\neg I'_n$ 
(in fact even from $\neg \Gamma'_n$) after Skolemization.  
From the assumption that the property holds for $I_n$ it follows that 
$G$ is a conjunction of atoms of
the form $c_{x} \rhd c_{y}, c_{x} \rhd t$, 
$f'(c_{x}) \rhd c_{y}$, $f'(c_{x}) \rhd s$, $g(c_y) 
\rhd c_z$, $g(c_y) \rhd s$ and $u \rhd v$,
where $t, s, u, v$ are constants and $c_x, c_y, c_z, ...$ are Skolem constants
introduced for the variables $x, y, z, ...$ (where $x, y, z$ are among
the variables in ${\overline x}$ universally quantified in $\Gamma'_n$).  

The analysis of the form of the candidate invariants obtained with
Algorithm 2
in the proof of Theorem~\ref{exhaustive-updates} shows that 
after Steps~1-3 of Algorithm~1, we obtain the following type of 
formulae: 

\smallskip
$ G'_{1(1)} \equiv  \!\!\!\!\!\!\!\!\!\! 
\displaystyle{\bigvee_{\stackrel{i_1, \dots, i_n \in  \{ 1, \dots, n_f \}}{D \subseteq {\sf est}(G)^2}}}
\!\!\!\!\!\!\!\!\!\! ~~~~~\Bigg( \displaystyle{\bigwedge_{p = 1}^k \phi^0_{i_p}(d_p)  \wedge \bigwedge_{(d_1, d_2) \in D} 
\!\!\!\! d_1 \not\approx d_2} \wedge \\
\exists x_{f'd_1}, \dots x_{f'd_n} \bigg( \displaystyle{\bigwedge_{p = 1}^k C^0_{i_p}(d_p, x_{f'd_p})  \wedge  \!\!\!\!\!\!\!\!\!\!\!\!\!\!  \bigwedge_{(f'(d_1),  f'(d_2)) \in {\sf est}(G)^2 \backslash D} \!\!\!\!\!\!\!\!\!\!\!\!\!\!
  x_{f'd_1} \approx x_{f'd_2} }  \wedge G^0_0({\overline d}, {\overline c},
{\overline c_{gc}}, {\overline x_{f'd}} ) \bigg) \Bigg)$

\smallskip
\noindent where in this case ${\overline c}, {\overline d}$ are the Skolem constants
occurring in $G$, such that (with the notation in the proof of
Theorem~\ref{exhaustive-updates}), ${\sf est}(G) = \{ f'(d_1), \dots, f'(d_n) \}$ and
${\sf est}(G_1) = \{ g(c_1),
\dots, g(c_k) \}$. 
Quantifier elimination is applied to formulae of the form: 
$$\exists x_{fd_1}, \dots x_{fd_n} \bigg( \displaystyle{\bigwedge_{p = 1}^k C^0_{i_p}(d_p, x_{fd_p})  \wedge  \!\!\!\!\!\!\!\!\!\!\!\!\!\!  \bigwedge_{(d_1,  d_2) \in {\sf est}(G)^2 \backslash D} \!\!\!\!\!\!\!\!\!\!\!\!\!\!
  x_{fd_1} \approx x_{fd_2} }  \wedge G^0_0({\overline d}, {\overline c},
{\overline c_{gc}}, {\overline x_{fd}} ) \bigg)$$
Due to the assumptions on the literals that can occur in $\Gamma_n$,
$G$ and ${\sf Update}$, these formulae contain only literals of
the form $s \rhd t$, where $s$ and $t$ are constants of $\Sigma_P$, Skolem
constants in $\{ d_1, \dots, d_k, c_1, \dots, c_n \}$, constants of
the form $c_{fd}$ or $c_{gc}$, 
or variables of the form $x_{f'd}$ which need to be eliminated
(where $\rhd \in \{ \leq, <, \geq, >, =
\}$). After quantifier elimination we obtain a disjunction
$G''_{1(1)}$ of conjunctions of literals of the form $s \rhd t$, where
$s$ and $t$ are constants of $\Sigma_P$, Skolem
constants in $\{ d_1, \dots, d_k, c_1, \dots, c_n \}$, or constants of
the form $c_{fd}$ or $c_{gc}$; after  Step 4 we obtain a disjunction
of conjunction of literals of the form 
$s \rhd t$, where $s$ and $t$ are constants of $\Sigma_P$, Skolem
constants in $\{ d_1, \dots, d_k, c_1, \dots, c_n \}$, or terms of the
form $f(c)$, $g(c)$, with $c \in \{ d_1, \dots, d_k, c_1, \dots, c_n
\}$. After replacing the Skolem constants $c_i, d_i$ again with
variables $x_i, y_i$ and negating in Step 5 we obtain a conjunction of 
clauses, each clause containing at most as many universally quantified 
variables as $\Gamma_n$, say $\{ x_1, \dots, x_m \}$, and only terms of the form 
$s \rhd t$, where $s$ and $t$ are constants in $\Sigma_P$, 
variables  $x_i$, or terms of the form $f(x_i), g(x_i)$ with $f \in
\Sigma, g \in \Sigma_P$. 
Note that under the assumptions we made the number of variables 
in the newly generated clauses does not grow. For a fixed set of 
variables $\{ x_1, \dots, x_m \}$ there are only finitely many terms
of the form above. Thus, after a finite number of steps no new
formulae can be generated and the algorithm terminates. 
\QED

\subsection{Termination vs. non-termination}
The class of formulae for which termination can be guaranteed is
relatively restricted. Even for transition systems describing programs
handling integers termination cannot be guaranteed, as already pointed
out in \cite{Dillig}.
While testing our method, we noticed that in some cases in which 
Algorithm 2 does not terminate, if we eliminate more symbols 
(thus restricting the language of the formula that strengthens the
property to be proved) we can obtain termination. 
We illustrate the ideas on two examples in which the theory 
is  linear integer arithmetic (so symbol elimination is 
simply quantifier elimination), then we discuss examples 
referring combinations of linear arithmetic with additional 
function symbols. 

\subsubsection{Constraints in linear integer arithmetic}

We consider the following examples. 

\begin{ex}
Consider the transition system $T = (\Sigma_S, {\sf Init}, {\sf
  Update})$ with $\Sigma_S = \Sigma_0 \cup \{x, y, z \}$, where
$\Sigma_0$ is the signature of linear integer arithmetic and $x, y, z$
are 0-ary function symbols, 
${\sf Init} = (x \approx 0 \wedge y \approx 0 \wedge  z  \approx 0)$ and
${\sf Update} = (x \leq N \rightarrow (x' \approx x+1 \wedge y' \approx y+1
\wedge z' \approx z+ x'-y'))$. 
Let $\Psi = (z \leq 0)$. 
$\Psi$ is true in the initial state; it is invariant under transitions 
iff 
$$z \leq 0 \wedge x \leq N \wedge (x' \approx x+1 \wedge y' \approx y+1 \wedge z'
\approx z+ x'-y') \wedge z' > 0$$
is unsatisfiable.  The formula is satisfiable. \\
After eliminating $x', y'$ and $z'$ we obtain 
$z \leq 0 \wedge x \leq N \wedge z + x - y > 0.$ \\
After negating this formula we obtain 
$\Gamma := ((z \leq 0 \wedge x \leq N) \rightarrow z \leq y - x )$. 
$\Psi \wedge \Gamma$ is not an inductive invariant. 
If we iterate the procedure we would obtain: 
$$\begin{array}{ll}
z \leq 0 & \wedge ~( x \leq N \rightarrow z \leq y - x) \\
& \wedge ~(x + 1 \leq N \rightarrow z \leq 2 y - 2 x) \\
& \dots
\end{array}$$
The process would not terminate. 

\medskip
\noindent We now consider an alternative procedure in which at the beginning in
addition to $x', y', z'$ we also eliminate $z$. We then obtain 
$x \leq N \wedge y - x < 0;$
after negation we obtain 
$$\Gamma := (x \leq N \rightarrow x \leq y)$$
It can be checked that 
$z \leq 0 \wedge (x \leq N \rightarrow x \leq y)$ is an invariant.

\smallskip
\noindent This second solution 
would correspond to an adaptation of the algorithm, 
in which some of the variables/symbols would be eliminated 
and several possibilities would be tried in order to check 
in which case we have termination. 
\end{ex}

\begin{ex}
Consider the transition system $T = (\Sigma_S, {\sf Init}, {\sf
  Update})$ where $\Sigma_S = \Sigma_0 \cup \{x, y \}$, where
$\Sigma_0$ is the signature of linear integer arithmetic and $x$ and $y$
are 0-ary function symbols, and where 
${\sf Init} = (x \approx 0 \wedge y \approx 0)$ and 
${\sf Update} = (x \leq N \rightarrow (x' \approx x +y \wedge y'
\approx 2*y)$. 
Let $\Psi = (x \geq 0)$.
$\Psi$ is true in the initial state; it is invariant under transitions 
iff 
$$x \geq 0 \wedge x \leq N \wedge (x' \approx x+y \wedge y' \approx y+y) \wedge x'
< 0$$
is unsatisfiable. The formula is satisfiable, so $\Psi$ is not an
inductive invariant. \\ 
After eliminating $x'$ and $y'$  we obtain 
$x \geq 0 \wedge x \leq N \wedge x + y < 0$.  \\
After negating we obtain 
$\Gamma := (x \geq 0 \wedge x \leq N) \rightarrow x + y \geq 0).$
$\Psi \wedge \Gamma$ is not an inductive invariant. 
If we iterate the procedure we obtain: 
$$\begin{array}{lll}
x \geq  0 & \wedge & (x \geq 0 \wedge x \leq N \rightarrow x + y \geq 0)
\\
&\wedge &  x \leq N  \wedge  
x+y \geq 0 \wedge x + y \leq N \wedge x + 3y < 0; 
\end{array}$$ 
after simplification:
$x \geq 0 \wedge x \leq N \wedge x + y \geq 0 \wedge x + y \leq N
\wedge x + 3y < 0;$ \\
after negation: $(x \geq 0 \wedge x \leq N \wedge x + y \geq 0 \wedge x + y \leq N) 
\rightarrow x + 3y \geq 0.$

\noindent This is not an invariant; with every iteration the coefficient of $y$
becomes larger; Algorithm 2 does not terminate.

\medskip 
\noindent We can choose to eliminate in addition to $x'$ and $y'$ also
$x$. 
After eliminating $x$ we obtain $y < 0$. 
After negation we obtain $\Gamma := y \geq 0$. 
It can be checked that $x \geq 0  \wedge y \geq 0$ is an invariant. 

\smallskip
\noindent This second variant corresponds, again, to an adaptation of the algorithm, 
in which we would try whether termination can be achieved 
after also eliminating some of the non-primed variables.
\end{ex}
\subsubsection{More general theories}

We now illustrate the algorithm on an example involving 
arrays and updates. 
We present first a very inefficient solution in which Algorithm 2 is
used, then a solution in which we use the refinement of Algorithm 1 
for symbol elimination.

\begin{ex}
Consider the program in Fig.~\ref{figure-ex} (a variation of an example 
from \cite{Rybalchenko}). 
The task is to prove that if $a$ is an array with increasingly sorted elements, 
then the formula $\Psi := d_2 \geq a[d_1 + 1]$ is an
invariant of the program.  
$\Psi$ holds in the initial state since 
$d_2 = a[4] = a[3+1]  = a[d_1 + 1] $. 
In order to show that $\Psi$ is an inductive invariant of the while
  loop, we would need to prove that the following formula is unsatisfiable: 
$$\begin{array}{l} 
{\sf Sorted}(a) \wedge d_1' \approx a[d_1 + 1] \wedge d_3' \approx
d_3/2 \wedge d_2 ' \approx a[d_2 + 1] + (1- d_3) \wedge d_2 \geq a[d_1 + 1] \wedge d_2' < a[d_1' + 1] ~~
\end{array}$$

\noindent where ${\sf Sorted}(a) := \forall i, j ( i \leq j 
\rightarrow a[i] \leq a[j])$. 
The updates change only constants and $\Psi$ is a ground formula. 
If ${\cal T} =  {\mathbb Z} \cup {\sf
  Sorted}(a)$ and  $G = d_2 \geq a[d_1 + 1]
\wedge d_1' \approx a[d_1 + 1] \wedge d_3' \approx d_3/2 \wedge d_2 ' \approx a[d_2 + 1]
+ (1- d_3) $, 
${\cal T} \wedge G$ is satisfiable iff the formula 
$\exists d_1 \exists d_2 \exists d_3 \exists d'_1 \exists d'_2 \exists
d'_3 G$ is valid w.r.t.\ ${\cal T}$. Satisfiability can be checked
using hierarchical reasoning. 

\medskip
\noindent {\bf Solution 1: Use Alg.\ 1 for symbol elimination.}
The (existentially) quantified variables $d_1', d_2'$ and $d_3'$ can
be eliminated. We obtain:  
$$\begin{array}{ll} 
{\sf Sorted}(a) \wedge & d_2 \geq a[d_1 + 1]  \wedge  a[d_2 + 1] + (1- d_3)  < a[a[d_1 + 1] + 1]
\end{array}$$
The axiom ${\sf Sorted}(a)$ defines a local theory extension $ {\mathbb Z} \subseteq {\mathbb Z} \cup {\sf
  Sorted}(a) = {\cal T}$; after
flattening of the ground part and instantiation of ${\sf Sorted}(a)$ we obtain: 

\smallskip
$\begin{array}{@{}ll}
G: & c_1 \approx a[d_1 + 1] \wedge d_2 \geq a[d_1 + 1] \wedge a[d_2 + 1] + (1- d_3)  < a[c_1 + 1]\\
{\sf Sorted}(a)[G]: &  d_1 + 1 \rhd d_2 + 1 \rightarrow a[d_1 + 1] \rhd a[d_2 + 1] \\
                           &  d_1 + 1 \rhd c_1 + 1 \rightarrow a[d_1 + 1] \rhd a[c_1 + 1] \\
                           & d_2 + 1 \rhd c_1 + 1 \rightarrow a[d_2 + 1] \rhd a[c_1 + 1], \rhd \in \{ \leq, \geq \} \\
\end{array}$ 

\smallskip
\noindent After purification in which the definitions ${\sf Def}:= \{
c_1 \approx a[d_1 + 1], c_2 \approx a[d_2 + 1], c_3 \approx a[c_1 + 1] \}$ are
introduced and further simplification we obtain: 

\smallskip
$\begin{array}{@{}ll}
G_0: & d_2 \geq c_1\wedge c_2 + (1- d_3)  < c_3 \\
{\sf Sorted}(a)[G]_0: & d_1 \rhd d_2 \rightarrow c_1 \rhd c_2 ~~
\wedge ~~ d_1 \rhd c_1  \rightarrow c_1 \rhd c_3 ~~ \wedge ~~ \\
& d_2 \rhd c_1 \rightarrow c_2 \rhd c_3,~~ \rhd \in \{ \leq, \geq \} \\
\end{array}$ 

\smallskip
\noindent 
By negating the formula above and universally
quantifying the constants we obtain a formula $\Gamma$ that 
can be used to strengthen the invariant. 
The process needs to be iterated, but the formulae are already
relatively large. 

\medskip
\noindent {\bf Solution 2: Use refinement of Alg.\ 1 for
  symbol elimination.}
Since ${\sf Sorted}(a)$ contains the parameter $a$ and no function
symbols occurring in this formula need to be eliminated, we do 
not consider any instances of ${\sf Sorted}(a)$ when doing quantifier elimination. 
The (existentially) quantified variables $d_1', d_2'$ and $d_3'$ can
be eliminated; we obtain: 
$$d_2 \geq a[d_1 + 1]  \wedge  a[d_2 + 1] + (1- d_3)  < a[a[d_1 + 1] + 1]$$
By negating the formula above and universally
quantifying the constants we obtain a formula 
$$\Gamma := d_2 \geq a[d_1 + 1]  \rightarrow  a[d_2 + 1] + (1- d_3)
\geq a[a[d_1 + 1] + 1]$$ 
that  can be used to strengthen the invariant to 
$$ \begin{array}{ll} 
& d_2 \geq a[d_1 + 1]   \wedge (d_2 \geq a[d_1 + 1]  \rightarrow  a[d_2 + 1] + (1- d_3)
\geq a[a[d_1 + 1] + 1]) \\
\equiv & d_2 \geq a[d_1 + 1] \wedge  a[d_2 + 1] + (1- d_3)
\geq a[a[d_1 + 1] + 1])
\end{array}$$ 
It is easy to see that with every iteration of Algorithm 2 we obtain
larger formulae. Algorithm 2 does not terminate.

\medskip
\noindent {\bf Solution 3: Eliminate also $a$.}
If we are looking, however, for a universal  invariant in a more
restricted language (for instance containing only the variables $d_1,
d_2$ and $d_3$), we can eliminate $c_1, c_2$ and obtain $d_3 > 1$. 
By negating this condition we obtain $d_3 \leq 1$. 
We obtain 
$d_2 \geq a[d_1 + 1] \wedge d_3 \leq 1$, which can
be proved to be a loop invariant. 
\label{ex-rybal-termination}
\end{ex}

\section{Tools for tests}
\label{implementation}

In this section we present the tools we used in our experiments and explain how we use them for the different steps of our method.

Essential to our method is the hierachical reduction of a formula
defining a chain of local theory extensions. As a tool for performing
the reduction we use H-PILoT \cite{hpilot} (which stands for Hierarchical Proving by Instantiation in Local Theory extensions). Given a formula or a set of clauses, H-PILoT performs the herarchical reduction to the base theory for any chain of theory extensions. Additionally, H-PILoT can be used for the proving task as well. For this it calls a prover for the base theory (for instance Z3, CVC4 or Yices) and outputs the answer of the selected prover. Note that the answer of H-PILoT can only be trusted, if the theory extensions are all local. Further information on H-PILoT can be found in \cite{hpilot}.

One very important procedure of our method is the quantifier
elimination. There are several tools available for using quantifier
elimination, for example Qepcad \cite{qepcad}, 
Mathematica \cite{mathematica} or 
Redlog \cite{redlog}. 
We mostly use Redlog, which is mainly for two reasons. Firstly, it's
one of the built-in provers for H-PILoT, and secondly, it can perform
QE for real closed fields as well as for Presburger arithmetic (QEPCAD
for example only supports QE for real numbers). 
% Further information on Redlog  can be found in \cite{redlog}.

\begin{figure}[tb]
	\centering
	\begin{minipage}{.9\linewidth}
        \includegraphics[width=0.8\textwidth]{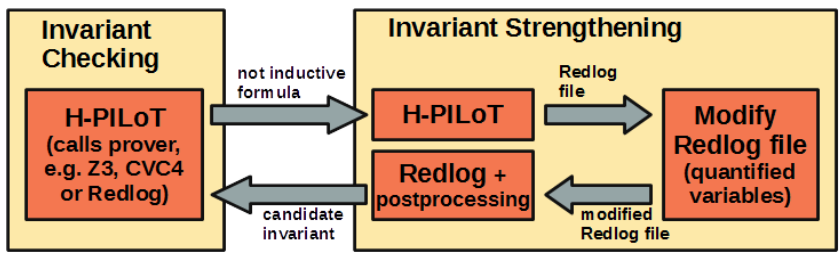}
	\caption{Current implementation of the algorithm}
	\label{fig:implementation}
	\end{minipage}
\end{figure}

Figure \ref{fig:implementation} schematically shows our current implementation. Our invariant generation method consists of two steps (that are possibly repeated). We first check whether a candidate invariant is indeed an invariant (invariant checking), and if it is not we generate a stronger invariant (invariant strengthening).

For the invariant checking we only need to use H-PILoT. It can do the reduction and call a prover. If the answer is "unsat", we know that our candidate invariant is an inductive invariant. If the answer is "sat", then it is not an inductive invariant and we have to apply the invariant strengthening. 

In the invariant strengthening step we first use H-PILoT to do the
reduction, but we don't let it call a prover. Instead we let it create
an input file for Redlog. If we would call Redlog directly from
H-PILoT, it would quantify all the variables. But since we only want
to eliminate some of the variables, we have to modify the Redlog file
accordingly (right now we have to do this manually, but we'd like to
add to H-PILoT in the future the option to select certain variables
that are to be eliminated by Redlog). We then use Redlog on the
modified Redlog file. As a postprocessing step the output formula
given by Redlog has to be negated. The conjunction of the old
candidate invariant and the formulae obtained this way 
is the new candidate invariant.

\section{Conclusion}
\label{conclusion}

\noindent We 
proposed a method for property-directed invariant 
generation and analyzed its properties.

\smallskip
\noindent 
We start from a given universal formula, describing a property 
of the data of the system. 
We can consider both properties on individual elements of an 
array (for instance:  $a[i] \leq b[j] + c[k]$ 
for fixed indices $i,j,k$) 
and "global properties'', for instance sortedness, or 
properties such as $\forall i (a[i] \leq b[i+1] - c[i-1] + d) $
(resp. e.g. $\forall i,j,k (a[i] \leq b[j] - c[k] - d)$ if we 
restrict to the array property fragment). 
These are properties which describe relationships which refer 
to the values of the variables or of the functions (e.g. arrays) 
at a given, fixed iteration in the execution of a loop. 
The invariants we generate have a similar form. 

\smallskip
\noindent 
Our results extend the results in \cite{Bradley12} and \cite{Dillig},  
as we consider more complex theories.
There are similarities to the method in \cite{Shoham16}, 
but our approach is different: The theories 
we analyze do not typically have the finite model property 
(required in \cite{Shoham17,Shoham16} where, if a 
counterexample $A$ to the inductiveness of a candidate
invariant $I$ is  found, a formula 
is added to $I$ to avoid finding the same counterexample 
again in the next iteration; to construct this formula the 
finite model property assumption is used). In our work 
we use the symbol elimination method in Alg.1 
to strengthen $I$; 
this should help to accelerate 
the procedure compared to the diagram-based approach. 
The decidability results in \cite{Shoham16} are 
presented in a general framework and rely on the 
well-foundedness of certain relations.  
In this paper we consider extensions of arithmetic (or other theories
allowing quantifier elimination) with 
additional function symbols; the theories we consider are not
guaranteed to have the finite model property.
For the situations in which we guarantee 
termination the abstract decidability or termination arguments 
in \cite{Shoham16}  might be difficult to check 
or might not hold (the arguments used for the case of pointers 
are not applicable). 
The algorithm proposed in \cite{Ghilardi10}  
for the theories of arrays uses a non-deterministic function 
{\sf ChooseCover} that returns a cover of a formula (as an approximation of 
the reachable states). It is proved that if the theory of 
elements is locally finite then 
for every universal formula $\Psi$, a universal inductive invariant $I$ 
strengthening $\Psi$ exists iff 
there exists a suitable {\sf ChooseCover} function for which the 
algorithm returns an inductive invariant strengthening $\Psi$.
In contrast to the algorithm proposed in \cite{Ghilardi10}, 
our algorithm is deterministic. 
To prove termination we show that the 
length of the quantifier prefix in the candidate 
invariants generated in every iteration does not 
grow; termination is then guaranteed if only finitely many atomic formulae 
formed with a fixed number of variables can be generated 
using quantifier elimination when applying the algorithm.

\smallskip
\noindent The methods used in \cite{Voronkov09,Voronkov09-fase,Voronkov-ijcar10}
and also in \cite{Gleiss18} 
often introduce a new argument to constants and functions symbols. 
If $a(f)$ is $n$, then an $n+1$-ary version of $f$ is used; 
$f(x_1, \dots, x_n, i)$ denotes the value of $f(x_1, \dots, x_n)$ at
iteration $i$. A major difference between our approach and the methods
for invariant generation used in 
\cite{Voronkov09,Voronkov09-fase,Voronkov-ijcar10} and \cite{Gleiss18}  
is that we do not use additional indices to refer to the values 
of variables at iteration steps and do not quantify over the 
iteration steps. However, an extension with quantification over iteration 
steps and possibilities of giving explicit solutions to 
at least simple types of recurrences seems to be feasible.

%%%% Out only in the long version %%% 
%% \smallskip
%% \noindent {\bf Towards an implementation.} 
%% We analyzed the applicability of our methods on several examples. 
%% In our tests, we used H-PILoT \cite{hpilot}  for the hierarchical
%% reduction (Step 1 in Alg.\ 1) and Redlog \cite{redlog} for quantifier
%% elimination (Step 3 in Alg.\ 1). 
%% % 
%% We selected the variables to be eliminated in Step 2 by analyzing
%% the problems and changed accordingly the redlog file given as output
%% by H-PILoT. 
%% Implementing this step is ongoing work.
%%%%%%%%%%%%
%% In Redlog we used both context OFSF 
%% (for quantifier elimination in real-closed fields) and context PASF
%% (for performing quantifier elimination in Presburger arithmetic)
%% and compared the results.  
%%%%%%%%%%%% 

\

\noindent 
{\bf Future work.} We here restricted to universally quantified invariants
and theories related to the array property fragment, but 
an extension to a framework using the notion 
of ``extended locality'' (cf.\
\cite{Sofronie-tacas08,Ihlemann-Sofronie-ijcar10}) seems
unproblematic. 
We plan to identify additional situations in which our
invariant generation method is correct, terminates resp. has low
complexity -- either by considering other theories or  
more general first-order properties.

\

\noindent {\bf Acknowledgments.} 
 We thank the reviewers of CADE-27 for their helpful comments.

\bibliographystyle{alpha}
\bibliography{bibliography2}
% \bibliography{bibliography2-short}

% \newpage 

\appendix
{

\section{Proof of Theorem~12} % \ref{exhaustive-updates}} 
\label{app:exhaustive-updates}

\noindent {\bf Theorem~\ref{exhaustive-updates}.}
Let $\Psi \in {\sf LocSafe}$ and ${\sf Update} = \bigvee_{f \in F}
{\sf Update}_f$ of the form discussed above. Assume that the clauses
in $\Psi$ and ${\sf Update}_f$ are flat and linear for all $f \in F$.
\footnote{We consider for the sake of simplicity only the case in
  which the functions $f \in F$ are unary.}  
Let $m$ be the maximal number of variables in a clause in $\Psi$. 
% , 
Assume that the only non-parametric functions which need to be
eliminated are the primed symbols $\{ f' \mid f \in F \}$ and that 
conditions ({\bf A1})--({\bf A5}) hold. 
Consider a variant of Algorithm~2, which uses for symbol elimination Algorithm~1 with the improvement 
in Theorem~\ref{symb-elim-simplif}.  Then for every step $n$, the clauses in the candidate invariant
$I_n$ obtained at step $n$ of Algorithm~2 are flat, the number of universally quantified variables in 
  every clause in $I_n$ is  $\leq m$. 

\

\noindent
{\em Proof:} 
By Theorem~\ref{symb-elim-simplif}, if the clauses $\K \cup I_n$ contain only function
symbols in $\Sigma_0 \cup \Sigma_P$, we need to apply Algorithm~1
to ${\sf Update}_f \wedge G$ only. 

\noindent 
To prepare the formula for symbol elimination we compute 
$G_1 = {\sf Update}_f[G] \cup G$ 
(the extension terms used in the instantiation are  {\sf est}(G) = $\{
f'(d_1), \dots, f'(d_k) \}$),
then instantiate also the terms starting with function symbols 
$g \in \Sigma_P \cup \Sigma$.  
Hence, we use the following set of ground terms: 
$T = {\sf est}(G) \cup {\sf est}(G_1) = \{ f'(d_1), \dots, f'(d_k) \} \cup \{ g(c) \mid
g \in \Sigma_P \cup \Sigma, g(c) \in {\sf est}(G_1) \}$ (a set of flat terms, in
which we isolated the terms starting with the function symbol $f$), 
By Lemma~\ref{lemma-DNF},  we have: 

\smallskip
\noindent $\begin{array}{@{}r@{}c@{}l} 
{\sf Update}_f[G] & := & \displaystyle{\bigwedge_{j = 1}^{k} \left( \bigwedge_{i = 1}^{n_f}
  (\phi_i(d_j) \rightarrow C_i(d_j, f'(d_j))) \right) 
\equiv
\bigwedge_{j = 1}^{k} \bigvee_{i = 1}^n (\phi_i(d_j) \wedge C_i(d_j,
f'(d_j))) } \\
& \equiv & \displaystyle{\bigvee_{i_1, \dots, i_k \in \{ 1, \dots,
    n_f\} }
\left(\bigwedge_{p = 1}^k \phi_{i_p}(d_p)  \wedge
\bigwedge_{p = 1}^k C_{i_p}(d_p, f'(d_p)) \right).} 
\end{array}$

\smallskip
\noindent We thus obtained a DNF with $(n_f)^k \leq (n_f)^m$ disjuncts, where $n_f$
(number of cases in the definition of $f$) and
$m$ (the maximal number of variables in $\K \cup I_n$) 
are constants depending on the description of the transition system. 
(We will show that this number does not increase in $\Gamma_{n+1}$.) 
Both $n_f$ and $m$ are typically small, in most cases 
% the definition of $f'$ is
% done by at most 3 disjoint cases, so 
$n_f \leq 3$. 

\noindent {\bf Step 1:} 
We introduce a constant $c_{f'd}$ for every term $f'(d) \in {\sf
  est}(G)$, replace  $f'(d)$ with $c_{f'd}$, add the
corresponding instances of the congruence axioms and obtain: 
$$\begin{array}{rcll}  
G_1 & = & G_0 \wedge & ({\sf Update}_f[G])_0 \wedge
\displaystyle{\bigwedge_{f'(d_1), f'(d_2) \in {\sf est}(G)} d_1 = d_2 \rightarrow
c_{f'd_1} = c_{f'd_2}}
\end{array}$$
We may compute a disjunctive normal form for the instances of congruence axioms or not; 
depending on this we obtain 
one of the equivalences~(\ref{equiv 1}) or (\ref{equiv 2}) below: 
\begin{eqnarray}
G_1 & \equiv G_0 \wedge \!\!\!\!\!\!\!\!\!\!
\displaystyle{\bigvee_{\stackrel{i_1, \dots, i_n \in \{ 1, \dots, n_f \}}{D \subseteq
    {\sf est}(G)^2}} } \bigg( & \displaystyle{\bigwedge_{p = 1}^k \phi_{i_p}(d_p)  \wedge
\bigwedge_{p = 1}^k C_{i_p}(d_p, c_{f'd_p}) } \wedge K_D({\sf Con}_0)
\bigg) \label{equiv 1} \\
 & \equiv G_0 \wedge \!\!\!\!\!\!\!\!\!\!
\displaystyle{\bigvee_{i_1, \dots, i_n \in  \{ 1, \dots, n_f \}} }
\bigg( & \displaystyle{\bigwedge_{p = 1}^k \phi_{i_p}(d_p)  \wedge
\bigwedge_{p = 1}^k C_{i_p}(d_p, c_{f'd_p}) } \wedge  {\sf Con}_0  \bigg)\label{equiv 2} 
\end{eqnarray}
In equivalence~(\ref{equiv 1}) we brought the formula to DNF. The
formulae $K_D({\sf Con})$ are the conjunctions which appear when
bringing the congruence axioms to DNF, based on a subset $D$ of 
${\sf  est}(G)^2$. In equivalence~(\ref{equiv 2}) we used distributivity and moved the
conjunction of all congruence axioms inside the conjunctive formulae
in the big disjunction.
 
\noindent Note that 
${\sf Con}_0 =  \displaystyle{\bigwedge_{f(d_1), {f_(d_2)} \in {\sf est}(G)}} (d_1 = d_2
\rightarrow c_{f'd_1} = c_{f'd_2})$  and every formula
$$K_D({\sf Con}_0) = \displaystyle{\bigwedge_{(f(d_1), f(d_2)) \in D} 
\!\!\!\!\!\! d_1 \neq d_2 \wedge \!\!\!\!\!\!\!\!\!\!\!\!\!\!\!\!\!\!\!  \bigwedge_{(f'(d_1),
    f'(d_2)) \in {\sf est}(G)^2 \backslash D} \!\!\!\!\!\!\!\!\!\!\!\!\!\!
  \!\!\!\! c_{f'd_1} = c_{f'd_2}} $$  
contain $k^2 \leq m^2$ conjunctions. 

In a second reduction we replace every term of the form
$g(d) \in {\sf est}(G_1)$, $g \in \Sigma_P$, with a new constant
$c_{gd}$. 
By Theorem~\ref{symb-elim-simplif} 
we do not need to add the corresponding congruence axioms. 
Using equivalence~(\ref{equiv 1}) and distributivity we obtain: 

\begin{eqnarray} 
G_{1(1)}^0 & \equiv &  \!\!\!\!\!\!\!\!\!\!\!\!\!\!\!\displaystyle{\bigvee_{\stackrel{i_1, \dots, i_k \in \{ 1, \dots, n_f \}}{D \subseteq {\sf est}(G)^2}}}
\!\!\!\!\! ~\bigg(\displaystyle{\bigwedge_{p = 1}^k \phi^0_{i_p}(d_p)  {\wedge}
\bigwedge_{p = 1}^k C^0_{i_p}(d_p, c_{f'd_p}) } {\wedge} K_D({\sf
Con}_0) {\wedge}  G^0_0({\overline d}, {\overline c}, {\overline c_{f'd}}, {\overline c_{gc}}) \bigg) \label{pos1} 
\end{eqnarray}
Using equivalence~(\ref{equiv 2}) and distributivity 
we obtain: 
\begin{eqnarray} 
G_{1(2)}^0 & \equiv & \!\!\!\!\!\!\!\!\!\! \displaystyle{\bigvee_{i_1,
    \dots, i_n \in \{ 1, \dots, n_f \}}}
\!\!\!\!\! ~~\bigg( \displaystyle{\bigwedge_{p = 1}^k \phi^0_{i_p}(d_p)}  {\wedge}
\displaystyle{\bigwedge_{p = 1}^k C^0_{i_p}(d_p, c_{f'd_p}) } {\wedge}
{\sf Con}_0 {\wedge} G^0_0({\overline d}, {\overline c}, {\overline c_{f'd}}, {\overline c_{gc}})\bigg) \label{pos2} 
\end{eqnarray}

\medskip 
\noindent where $\phi_j^0, C^0_j, G_0^0$ are obtained from $\phi_j,
C_j, G_0$
after replacing all terms of the form $g(c)$, $g \in \Sigma_P$ with
$c_{gc}$. 

\medskip
\noindent {\bf Step 2:} $f'$ is not a parameter; all the other
  function symbols are either parametric or in $\T_0$. 
We therefore replace the constants $c_{f'd}$ with variables $x_{f'd}$.  

\noindent {\bf Step 3:} 
Note that $\exists x_{f'd_1}, \dots x_{f'd_n} ~ G^0_{1(1)}(x_{f'd_1}, \dots, x_{f'd_n})$ is
  equivalent to:  

\begin{eqnarray} 
G'_{1(1)} & \equiv & \!\!\!\!\!\!\!\!\!\! 
\displaystyle{\bigvee_{\stackrel{i_1, \dots, i_n \in  \{ 1, \dots, n_f \}}{D \subseteq {\sf est}(G)^2}}}
\!\!\!\!\!\!\!\!\!\! ~~~~~\Bigg( \displaystyle{\bigwedge_{p = 1}^k \phi^0_{i_p}(d_p)  \wedge \bigwedge_{(f'(d_1), f'(d_2)) \in D} 
\!\!\!\! d_1 \neq d_2} \wedge  \label{pos-qe-1} \\
& & \exists x_{f'd_1}, \dots x_{f'd_n} \bigg( \displaystyle{\bigwedge_{p = 1}^k C^0_{i_p}(d_p, x_{f'd_p})  \wedge  \!\!\!\!\!\!\!\!\!\!\!\!\!\!  \bigwedge_{(f'(d_1),  f'(d_2)) \in {\sf est}(G)^2 \backslash D} \!\!\!\!\!\!\!\!\!\!\!\!\!\!
  x_{f'd_1} = x_{f'd_2} }  \wedge G^0_0({\overline d}, {\overline c},
{\overline c_{gc}}, {\overline x_{f'd}} ) \bigg) \Bigg)\nonumber 
\end{eqnarray}
Using equivalence~(\ref{equiv 2})
we obtain: 
\begin{eqnarray} 
G'_{1(2)} & \equiv & \!\!\!\!\!\!\!\!\!\! \displaystyle{\bigvee_{ i_1, \dots, i_n \in  \{ 1, \dots, n_f \}}}
\Bigg( \displaystyle{\bigwedge_{p = 1}^k \phi^0_{i_p}(d_p)}  \wedge
 \exists x_{f'd_1}, \dots x_{f'd_n} \bigg( \displaystyle{\bigwedge_{p =
     1}^k C^0_{i_p}(d_p, x_{f'd_p}) } \wedge   \label{pos-qe-2} \\
& & \displaystyle{\!\!\!\!\!\!\!\!\!\!\!\!\!\!  \bigwedge_{(f'(d_1), f('d_2)) \in {\sf est}(G)} 
\!\!\!\! (d_1 = d_2 \rightarrow x_{f'd_1} = x_{f'd_2})}  \wedge   G^0_0({\overline d}, {\overline c},
{\overline c_{gc}}, {\overline x_{f'd}} ) \bigg) \Bigg) \nonumber 
\end{eqnarray}
After quantifier elimination we obtain, using equivalence~(\ref{equiv 1}), a formula of the form: 
\begin{eqnarray} 
G''_{1(1)} & \equiv  & \!\!\!\!\!\!\!\!\!\! 
\displaystyle{\bigvee_{\stackrel{i_1, \dots, i_n \in \{ 1, \dots, n_f \}}{D \subseteq {\sf est}(G)^2}}}
\bigg( \displaystyle{\bigwedge_{p = 1}^k \phi^0_{i_p}(d_p)  \wedge \bigwedge_{(d_1, d_2) \in D} 
\!\!\!\! d_1 \neq d_2} \wedge D^1_{i_1,\dots,i_n,D}({\overline d},
{\overline c}, {\overline
c_{gc}}) \bigg)\label{pos-qe-1bis}
\end{eqnarray}
and using equivalence~(\ref{equiv 2}): 
\begin{eqnarray} 
G''_{1(2)} & \equiv & 
\!\!\!\!\!\!\!\!\!\!  
\displaystyle{\bigvee_{i_1, \dots, i_n \in  \{ 1, \dots, n_f \}}
\Bigg( \bigwedge_{p = 1}^k \phi^0_{i_p}(d_p)  \wedge  D^2_{i_1,\dots,i_n,D}({\overline d},
{\overline c}, {\overline
c_{gc}})  } \Bigg)  \label{pos-qe-2bis} 
\end{eqnarray}
where 
$D^1_{i_1,\dots, i_n,D}({\overline d}, {\overline c}, {\overline
c_{gc}})$ is obtained after QE from 

$\exists x_{f'd_1}, \dots x_{f'd_n}  \bigg(
\displaystyle{\bigwedge_{p = 1}^k C^0_{i_p}(d_p, x_{f'd_p})  \wedge
  \!\!\!\!\!\!\!\!\!\!\!\!\!\!  \bigwedge_{(f'(d_1),  f'(d_2)) \in {\sf
      est}(G)^2 \backslash D} \!\!\!\!\!\! x_{f'd_1} = x_{f'd_2} }  \wedge G^0_0({\overline d}, {\overline c}, {\overline c_{gc}}, {\overline x_{f'd}})  \bigg)$

\noindent and $D^2_{i_1,\dots,i_n,D}({\overline d}, {\overline c},  {\overline
c_{gc}})$ is obtained after QE from  

\noindent $\exists x_{f'd_1}, \dots x_{f'd_n}\bigg( \displaystyle{\bigwedge_{p = 1}^k C^0_{i_p}(d_p, x_{f'd_p}) } \wedge  \displaystyle{\!\!\!\!\!\!\!\!\!\!\!\!\!\!  \bigwedge_{(f'(d_1), f'(d_2)) \in {\sf est}(G)} 
\!\!\!\!\!\!\!\! (d_1 = d_2 \rightarrow x_{f'd_1} = x_{f'd_2})}  \wedge G^0_0({\overline d}, {\overline c}, {\overline c_{gc}}, {\overline x_{f'd}})  \bigg).$

\

\noindent {\bf Step 4:} We replace back in the formula obtained this
way all constants $c_{gd}$, $g \in \Sigma_P, g(c) \in {\sf est}(G_1)$
with the terms $g(c)$.  
We obtain therefore, if starting from formula~(\ref{equiv 1}): 

$ \begin{array}{ll} 
\Gamma_{2(1)}({\overline d}, {\overline c}) = 
\!\!\!\!\!\!\!\!\!\! \displaystyle{\bigvee_{ \stackrel{i_1, \dots, i_n \in  \{ 1, \dots, n_f \}}{D \subseteq {\sf est}(G)^2}}}
\Bigg( & \bigg( \displaystyle{\bigwedge_{p = 1}^k \phi_{i_p}(d_p)}  \wedge \displaystyle{\bigwedge_{(d_1, d_2) \in D}
\!\!\!\! d_1 \neq d_2 \wedge G^g_0({\overline d}, {\overline c}, {\overline g(c)}) 
\bigg)} \wedge \\
& D^1_{i_1,\dots,i_n,D}({\overline d}, {\overline c}, {\overline
g(c)}) \Bigg)
\end{array}$

\noindent and if starting from formula~(\ref{equiv 2}): 

\noindent $\begin{array}{ll} 
\Gamma_{2(2)}({\overline d}, {\overline c}) = 
\!\!\!\!\!\!\!\!\!\! \displaystyle{\bigvee_{ i_1, \dots, i_n \in  \{ 1, \dots, n_f \}} \bigvee_{D \subseteq {\sf est}(G)^2}}
\Bigg( & \bigg( \displaystyle{\bigwedge_{p = 1}^k \phi_{i_p}(d_p)}
\wedge G^g_0({\overline d}, {\overline c}, {\overline g(c)}) \bigg) \wedge D^2_{i_1,\dots,i_n,D}({\overline d}, {\overline c}, {\overline
g(c)})\Bigg)
\end{array}$

\

\noindent {\bf Step 5:} We negate $\exists {\overline y}
\Gamma_{2(1)}({\overline y})$ and obtain: 

$$\begin{array}{ll} 
 \forall y \Bigg[\!\!\!\!\!\!\!\!\!\! \displaystyle{\bigwedge_{i_1, \dots,
      i_n \in \{ 1, \dots, n_f \}} \bigwedge_{D \subseteq {\sf est}(G)^2}}
\Bigg( & \neg  \bigg( \displaystyle{\bigwedge_{p = 1}^k \phi_{i_p}(y_p)}  \wedge \!\!\!\! \displaystyle{\bigwedge_{(f(d_1), f(d_2)) \in D}}
\!\!\!\! y_1 \neq y_2 \wedge G^g_0({\overline y}, {\overline g(y)}) \bigg) \wedge \\
& \neg D^1_{i_1,\dots,i_n,D}({\overline y},
{\overline g(y)}) \Bigg) \Bigg]
 \end{array}$$
which is equivalent with: 
 $$\begin{array}{ll} 
 \forall y \Bigg[\!\!\!\!\!\!\!\!\!\! \displaystyle{\bigwedge_{ i_1, \dots, i_n \in \{ 1, \dots, n_f \}}\bigwedge_{D \subseteq {\sf est}(G)^2}}
\Bigg( & \bigg( \displaystyle{\bigwedge_{p = 1}^k \phi_{i_p}(y_p)}  \wedge \displaystyle{\bigwedge_{(f(d_1), f(d_2)) \in D}}
\!\!\!\! y_1 \neq y_2 \wedge G^g_0({\overline y}, {\overline g(y)}) \bigg) \wedge \\
& \rightarrow \neg D^1_{i_1,\dots,i_n,D}({\overline y},
{\overline g(y)}) \Bigg) \Bigg]
 \end{array}$$

Similarly for $\exists {\overline y} \Gamma_{2(2)}({\overline y})$.

\medskip
\noindent
{\bf Complexity:} We analyze the complexity of the transformations
presented above. We show how the size of the formula changes depending
on $n_f$ and $m$ (since $k \leq m$ and $m$ is a constant of the
system). 
After the instantiation we first have a conjunction of $k * n_f \leq m * n_f$ implications. 
After the transformation to DNF we obtain a disjunction of at most
$n_f^m$ conjunctions, each of length  at most $m * l_f$, where $l_f$ is the
maximal length of a rule in ${\sf Update}_f$.  
After applying the first reduction in Step 1 
we obtain a disjunction of $n_f^m * 2^{m^2}$ conjunctions each of
length at most $(m * l_f + m^2 + |G|)$  (with
transformation (1)) resp.\ of $n_f$ formulae, each of length 
at most $(m * l_f + |G| + |{\sf Con}_0|)$ (with transformation
(2)).  Note that $m * l_f$ and $m^2$ are constants of the system. 
In Step 2 the size of the formulae does not change. 
After quantifier elimination in Step 3 the size of the conjuncts may
grow: for every variable which is eliminated, the size of the formula
might grow quadratically. 
The last two steps of the algorithm do not change the size of the
formula.

\end{document}